\documentclass[a4paper,fleqn]{cas-dc}
\usepackage{hyperref}
\hypersetup{colorlinks,allcolors=black}
\usepackage[utf8]{inputenc}
\usepackage[numbers]{natbib}
\usepackage{amsmath,amssymb,amsfonts}
\usepackage{algorithmic}
\usepackage{algorithm}
\usepackage{array}
\usepackage{graphicx}
\usepackage{xcolor}
\usepackage{multirow}
\usepackage{placeins}
\usepackage{tabularx}
\usepackage{textcomp}
\usepackage{stfloats}
\usepackage{url}
\usepackage{verbatim}
\usepackage{subcaption}
\usepackage{booktabs} 

\usepackage{wrapfig} 
\usepackage{flushend} 
\def\tsc#1{\csdef{#1}{\textsc{\lowercase{#1}}\xspace}}
\tsc{WGM}
\tsc{QE}
\tsc{EP}
\tsc{PMS}
\tsc{BEC}
\tsc{DE}

\begin{document}
\let\WriteBookmarks\relax
\def\floatpagepagefraction{1}
\def\textpagefraction{.001}

\shorttitle{CodeBC: A More Secure Large Language Model for Smart Contract Code Generation in Blockchain}

\shortauthors{L. Wang et~al.}

\title [mode = title]{CodeBC: A More Secure Large Language Model for Smart Contract Code Generation in Blockchain}                      



%
\author[1,2,3]{Lingxiang Wang}[type=editor,
                        orcid=0009-0001-0160-5701]


\ead{wanglingxiang@buaa.edu.cn}



\affiliation[1]{organization={State Key Laboratory of Complex \& Critical Software Environment, Beihang University},
    city={Beijing},
    postcode={100190}, 
    country={China}}

\affiliation[2]{organization={School of Artificial Intelligence, Beihang University},
    city={Beijing},
    postcode={100190}, 
    country={China}}

\affiliation[3]{organization={Beijing Advanced Innovation Center for Future Blockchain and Privacy Computing},
    city={Beijing},
    postcode={100190}, 
    country={China}}

\author[1,2,3]{Hainan Zhang}[style=chinese]
\cormark[1]
\ead{zhanghainan@buaa.edu.cn}

\author[2,3]{Qinnan Zhang}[style=chinese]
\ead{zhangqn@buaa.edu.cn}

\author[1,2,3]{Ziwei Wang}[style=chinese]
\ead{wangziwei26@buaa.edu.cn}



\author[3,4]{Hongwei Zheng}[%
   ]
\ead{zhenghongwei2024@163.com}

\author[3,4]{Jin Dong}[%
   ]
\cormark[1]
\ead{dongjin@baec.org.cn}

\author[1,2,3]{Zhiming Zheng}
\ead{zzheng@pku.edu.cn}

\affiliation[4]{organization={Beijing Academy of Blockchain and Edge Computing, BABEC},
    city={Beijing},
    postcode={100086}, 
    country={China}}

\cortext[cor1]{Corresponding author}


\nonumnote{This work was funded by the National Natural Science Foundation of China (NSFC) under Grants No. 62406013, the Beijing Advanced Innovation Center for Future Blockchain and Privacy Computing, the State Key Laboratory of Complex \& Critical Software Environment and the Fundamental Research Funds for the Central Universities.
  }

\begin{abstract}
Large language models (LLMs) excel at generating code from natural language instructions, yet they often lack an understanding of security vulnerabilities. This limitation makes it difficult for LLMs to avoid security risks in generated code, particularly in high-security programming tasks such as smart contract development for blockchain. Researchers have attempted to enhance the vulnerability awareness of these models by training them to differentiate between vulnerable and fixed code snippets. However, this approach relies heavily on manually labeled vulnerability data, which is only available for popular languages like Python and C++. For low-resource languages like Solidity, used in smart contracts, large-scale annotated datasets are scarce and difficult to obtain.
To address this challenge, We introduce CodeBC, a code generation model specifically designed for generating secure smart contracts in blockchain. CodeBC employs a three-stage fine-tuning approach based on CodeLlama, distinguishing itself from previous methods by not relying on pairwise vulnerability location annotations. Instead, it leverages vulnerability and security tags to teach the model the differences between vulnerable and secure code. During the inference phase, the model leverages security tags to generate secure and robust code. 
Experimental results demonstrate that CodeBC outperforms baseline models in terms of BLEU, CodeBLEU, and compilation pass rates, while significantly reducing vulnerability rates. These findings validate the effectiveness and cost-efficiency of our three-stage fine-tuning strategy, making CodeBC a promising solution for generating secure smart contract code.
\end{abstract}



\begin{NoHyper}
\maketitle
\end{NoHyper}
\section{INTRODUCTION}
\begin{figure}[]
    \centering
    \includegraphics[width=\columnwidth]{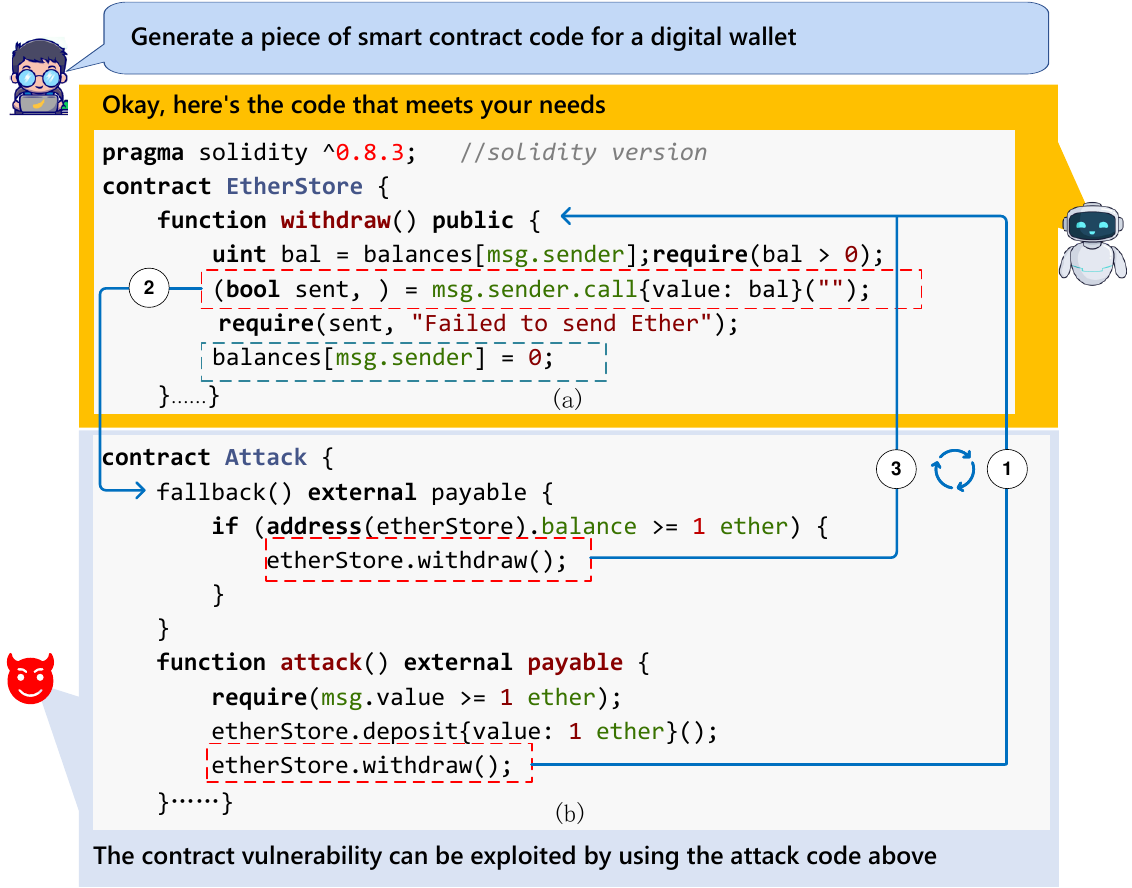}
    \captionsetup{justification=justified, singlelinecheck=false, labelfont=bf}
    \caption{An example of generative code with smart contract reentrancy attack. The process of reentrancy attack by contract(b) on contract(a):\text{\textcircled{1}} The attacker initiates a withdrawal request; \text{\textcircled{2}} EtherStore responds and automatically calls the \textit{fallback()} function. \text{\textcircled{3}} The attacker re-initiates the withdrawal request. Since the account balance has not been cleared, the loop can continue.}
    \label{fig:contract}
    \vspace{-1.5em}
\end{figure}

Large language models have excelled in instruction-based code generation tasks. As show in the upper part of Figure~\ref{fig:contract}, LLMs can effectively understand user instructions and generate code that meets these requirements~\citep{alphacode}. This advancement marks a major breakthrough in software engineering. It can streamline the software development process and reduce human error.
However, most research focuses on enhancing the functional completeness of code generation. Little attention is given to the security of generated code. This includes identifying potential vulnerabilities or assessing whether the code could cause harm, such as information leaks or financial losses. This oversight limits their application in vulnerability-sensitive domains. Blockchain is one such domain, where code security is essential for users and communities.
Blockchain is a decentralized and distributed ledger database. It records transactions across multiple computers in a secure, transparent, and tamper-resistant manner. To ensure transaction security and transparency, developers must pre-define transaction rules using smart contract code. The code should execute automatically without intermediaries when invoked.
The security of smart contract code is critical. If the code contains vulnerabilities, malicious actors might exploit them. This could result in significant financial losses for users. The lower part of Figure~\ref{fig:contract} illustrates an example of code generated by LLMs with a reentry attack\footnote{
A typical withdrawal operation should clear the account balance before processing the withdrawal. However, the withdrawal function defined in contract(a) does the opposite. So, when an attacker's contract(b) continuously calls contract(a), it results in erroneous behavior of continuously sending virtual currency.}. The example shown in the figure is intended solely for illustrative purposes and may appear overly simplistic\footnote{
Consequently, when using the prompt shown in the figure to generate smart contract code with current popular models, the reentrancy attack vulnerability illustrated in the figure will not arise.} 
More comprehensive prompts, including more detailed examples of generating code with vulnerabilities, can be found in our open-source Blockchain-Humaneval dataset\footnote{https://github.com/wanglingxiang1298/CodeBC/tree/master/dataset\\/result\_dataset/solidity/BlockChain-Humaneval}.
The experiments carried out by Hammond Pearce et al. also emphasize that the code generated by certain large language models exhibits security deficiencies~\citep{Pearce_Ahmad_Tan_Dolan-Gavitt_Karri_2022}.
If high-quality code can be generated by LLMs for blockchain, it can not only enhance developer efficiency but also reduce security risk and prevent economic losses. Therefore, how to reduce vulnerabilities of code generated by LLMs has become a key issue in applying them to the blockchain smart contract.

Researchers~\citep{he2023large} have attempted to enhance the vulnerability detection capabilities of code generation models. They train the models to differentiate between vulnerable code and its corresponding fixed code. This approach controls code generation through prefix-tuning and applies contrastive learning between the original vulnerable code and its manually fixed version.
However, this method is limited to popular programming languages like Python and C++, which have abundant annotated pairwise vulnerability datasets. In contrast, low-resource languages, such as Solidity in smart contracts, face a different situation. While large-scale open-source datasets are available for vulnerability detection, they only indicate whether a contract contains vulnerabilities, without providing details on how to fix them.
Since LLMs require substantial data to perform well, a lower-cost and more effective method is needed. This method should inject vulnerability knowledge of smart contracts into LLMs to enhance the security of generated code in the blockchain domain.

To achieve this goal, we propose CodeBC, a more secure code generation model for blockchain. It employs a customized three-stage fine-tuning approach to enhance the model's understanding of smart contract code, vulnerability knowledge, and human instructions.Specifically, in the first stage, we employ code infilling task to strengthen the model's comprehension of smart contract code, which randomly mask some lines of code as inputs and use the masked code segments as outputs.
In the second stage, we inject vulnerability knowledge into the model through a vulnerability detection task. This stage extracts code from open-source vulnerability detection datasets and security repositories as input, with vulnerability and security tags as outputs.
In the third stage, we apply a tag-guided instruction fine-tuning method to improve the model's ability to follow human instructions. We concatenate human instructions with the security or vulnerability tags from the second stage as input, and use the corresponding code as output.
During inference, we concatenate human instructions and security tag to control the model to generate more secure code.


To better evaluate whether the generated code meets human instruction requirements, we construct a dataset called Blockchain-HumanEval. This dataset is based on the widely used open-source repository OpenZeppelin, with manual instruction annotations according to its functionalities.
The experimental results show that, compared to baseline models, our model achieves significant improvements in BLEU, CodeBLEU scores, compilation pass rate, and vulnerability-free rate. Compared to the baseline SVEN, which requires pairwise vulnerability annotations, our approach delivers comparable performance for Python and C/C++. This demonstrates that our low-cost method is both simple and effective. We will publish models and datasets when this paper is accepted\footnote {https://github.com/wanglingxiang1298/CodeBC/tree/master}.

The innovations of this paper are as follows:
\begin{itemize}
    \item We introduce an instruction-based code generation model into blockchain field, which can enhance developer efficiency but also reduce security risk, fostering interdisciplinary research.

    \item We propose a customized three-stage fine-tuning strategy: code infilling to enhance adaptability to blockchain domain, vulnerability detection to inject the vulnerability knowledge, and tags-guided approach to improve comprehension of human instructions. 
    
    \item We construct the first human evaluation dataset for instruction-based smart contract code generation task, and experimental results demonstrate the effectiveness and low-cost of our method.
\end{itemize}

The paper is organized as follows. In Section~\ref{background}, we review some essential background. Section~\ref{codebc model} introduces the design of CodeBC, including the details of three-stage fine-tuning strategy. Section~\ref{exp setup} introduces some empirical settings. Experiment results and ablation results on BlockChian-HumanEval datasets are demonstrated in Section~\ref{exp result}. In Section~\ref{dis}, we evaluate the effectiveness of three-stage fine-tuning on various tasks. 
Finally, Section~\ref{related work} compares our approach against some related work, before Section~\ref{conclusion} concludes.
\section{BACKGROUND}\label{background}
\subsection{Vulnerability Types of Smart Contract}
Here are some typical types of smart contract vulnerabilities~\footnote{https://huggingface.co/datasets/mwritescode/slither-audited-smart-contracts}.

\textbf{Reentrancy}(RE): Smart contracts can interact with external contracts or accounts. However, improper design may introduce reentrancy vulnerabilities. These vulnerabilities allow attackers to repeatedly call a function during its execution, enabling malicious actions. A well-known example is ``The DAO" attack, which resulted in nearly \$50 million in losses.

\textbf{Access Control}(AC): Smart contracts rely on permission controls and condition checks to restrict access to functions and data. Weak access controls may allow attackers to gain unauthorized privileges, resulting in data leaks, financial losses, or contract disruptions. On the other hand, overly strict controls can lock assets and cause economic losses. A notable example is the 2017 Parity wallet attack, where attackers bypassed access controls and stole approximately \$30 million.

\textbf{Arithmetic}(AR): Integer overflow and underflow are critical vulnerabilities in smart contracts due to the frequent use of unsigned integers and simple types. Overflow can transform benign contracts into tools for theft or denial-of-service attacks. A well-known example is the 2018 BEC attack. In this case, attackers flooded the exchange with tokens, causing the BEC price to drop to zero and resulting in significant economic losses.

\textbf{Unchecked Low Level Calls}(ULLC): Solidity's low-level functions, such as call(), delegatecall(), and staticcall(), return a boolean value of false on failure instead of propagating or reverting errors. If not properly checked, these functions may allow attackers to execute unauthorized operations, manipulate contract states, or steal funds.

\textbf{Denial of Service}(DoS): DoS attacks pose a major threat to blockchain systems, including Ethereum. These attacks involve irreversible malicious operations or excessive resource consumption. In Ethereum, each request consumes gas, and execution stops once the gas limit is reached. DoS attacks can disrupt contract execution and lead to substantial token and gas consumption.

\textbf{Bad Randomness}(BR): BR is a challenge in modern computing systems, particularly in open blockchain networks like Ethereum. Generating random numbers in smart contracts is difficult because on-chain data is publicly accessible. If not carefully managed, this can be exploited for cheating.

\textbf{Front Running}(FR): In Ethereum, transactions are confirmed based on the fees paid to miners. If an attacker obtains transaction information in advance, they can outbid the original user by increasing the fee, completing the operation first and causing losses to the user.

\textbf{Time Manipulation}(TM): In Solidity, using timestamp can introduce vulnerabilities because it is controlled by miners. Attackers, typically miners, can manipulate the timestamp to alter the outcome of contracts that depend on it, achieving their desired results.

\subsection{Language Models}
\begin{figure}[!t]
    \centering
    \includegraphics[width=\columnwidth]{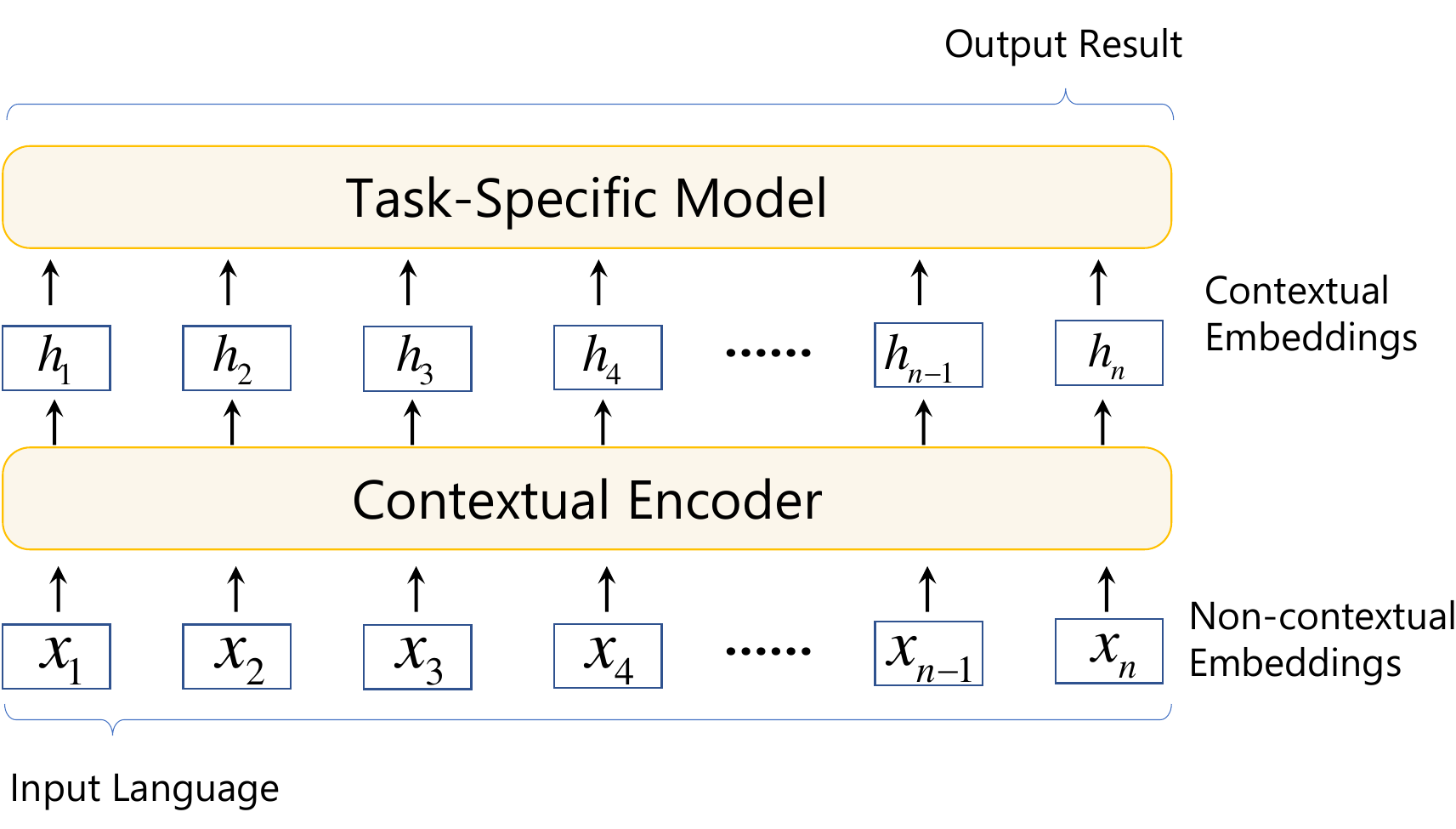}
    \captionsetup{justification=justified, singlelinecheck=false, labelfont=bf}
    \caption{General architecture of a language model}
    \label{fig:lm}
    \vspace{-1em}
\end{figure}

The rapid development of language models, particularly with the introduction of the transformer architecture~\citep{vaswani2017attention}, has significantly improved language representations and the use of contextual information. Numerous pre-trained models for general language representation, such as BERT~\citep{Bert}, GPT~\citep{GPT}, and Llama~\citep{Llama}, have since been proposed. These models have shown excellent performance in various natural language processing tasks, including programming and question answering.
Figure \ref{fig:lm} illustrates the general architecture of a language model. The input is processed through Non-contextual embeddings and Contextual embeddings to obtain the encoded state, which is then provided to a task-specific model to yield the output result.
In Non-contextual embeddings, discrete linguistic symbols are mapped into a distributed embedding space. For a word \( \text{i} \) in the vocabulary \( \mathcal{V} \), it is mapped to a vector \( \mathbf{x}_i \in \mathbb{R}^{D_e} \) in the lookup table \( \mathbf{E} \in \mathbb{R}^{D_{e} \times |\mathcal{V}|} \), where \( D_e \) is the dimension of the embedding. Subsequently, to further understand contextual information, a neural encoder is used to encode the word embedding into a context-based word embedding. Specifically, given a sequence of word embeddings \( [x_1, x_2, \ldots, x_n] \) where \( x_i \) is a token after word embedding and \( x_i \in \mathcal{V} \), the contextual representation of \( x_i \) depends on the entire text segment.
\begin{equation}[\mathbf{h}_1,\mathbf{h}_2,\ldots,\mathbf{h}_T]=f_{\mathrm{enc}}\left(x_1,x_2,\ldots,x_T\right),
\end{equation}
where \(f_{\mathrm{enc}}(\cdot)\) represents the neural encoder, and \(h_i\) is the context-based embedding for the token \(x_i\).
The purpose of encoding language is to establish a conditional distribution that continuously predicts the next token based on the known sequence of tokens. Given a text sequence \(x_{1:N}=[x_{1},x_{2},\ldots,x_{n}]\), its joint probability \(p\left(x_{1:N}\right)\) can be factored as follows:
\begin{equation}
p\left(x_{1:N}\right)=\prod_{i=1}^{y}p\left(x_{i}\mid x_{0:i-1}\right),
\end{equation}
where \(x_0\) is a special token representing the start of the sequence. The conditional probability \(p\left(x_{i}\mid x_{0:i-1}\right)\) can be modeled to predict the probability distribution of a word given the language context \(x_{0:i-1}\). The context \(x_{0:i-1}\) can be modeled through the neural encoder \(f_{\mathrm{enc}}(\cdot)\), and thus the conditional probability can be expressed:
\begin{equation}
p\left(x_i|x_{0:i-1}\right)=g_{\mathrm{LM}}\left(f_{\mathrm{enc}}\left(x_{0:i-1}\right)\right),
\end{equation}
where \(g_{\mathrm{LM}}\) represents the prediction layer.
Hence, this paper constructs structured information based on vulnerability tags, which better serves as a reference for model inference by incorporating vulnerability information.

\section{CODEBC MODEL}\label{codebc model}
\begin{figure*}[]
    \centering
    \includegraphics[width=\textwidth]{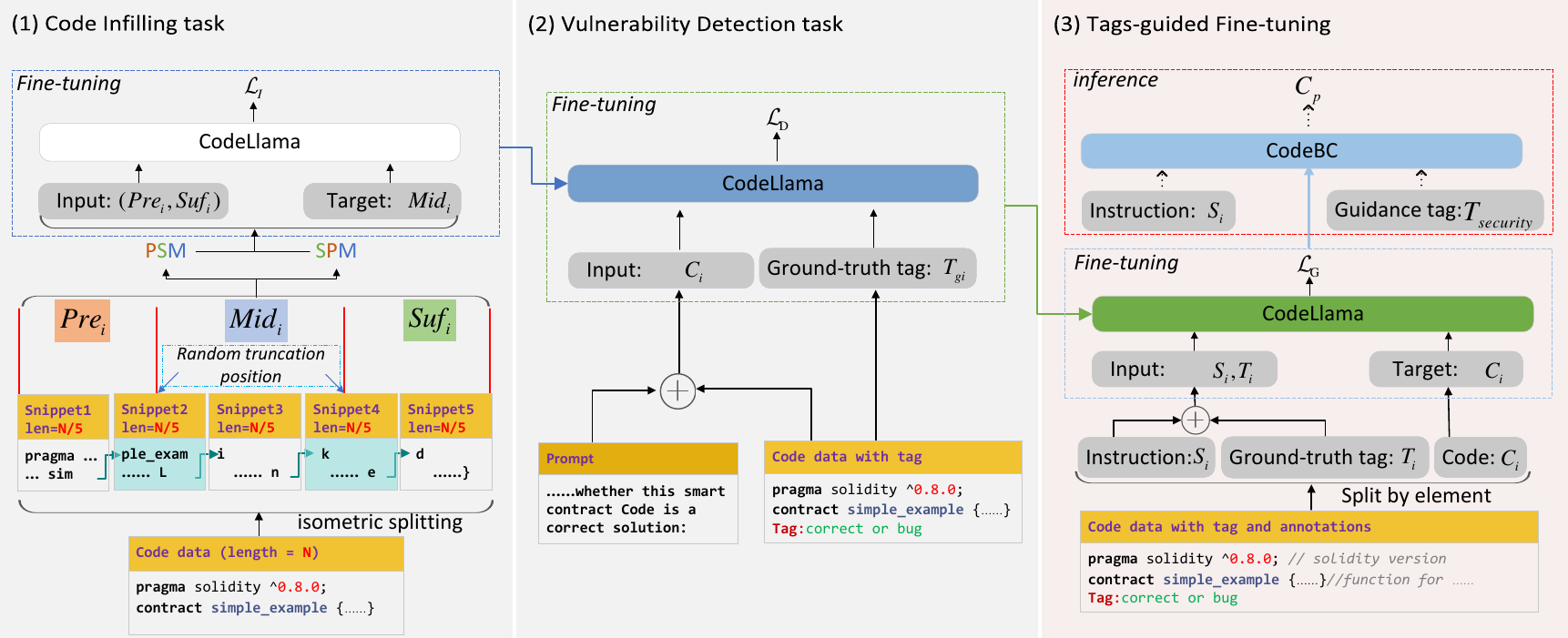}
    \captionsetup{justification=justified, singlelinecheck=false, labelfont=bf}
    \caption{An illustration of the CodeBC model. $(1)$ First fine-tuned the base pre-trained language model on smart contract code. $(2)$ Continued to train the model with the goal of vulnerability detection task on the model with the first smart contract code generation capability. $(3)$ Finally, Tags-guided Instruction Fine-tuning was performed on the models, aligning the vulnerability detection tags in stages $(2)$.}
    \label{fig:framework}
    \vspace{-1em}
\end{figure*}
\subsection{Task Definition}

The objective of the instruction-based smart contract code generation task is to produce smart contract code that meets the user's requirements. For the task of secure contract code generation, the goal is to ensure the model generates contract code that meets the user's requirements and avoids vulnerabilities. To achieve this goal, the model needs to have a sufficient reserve of smart contract code knowledge, understand contract vulnerabilities, and then be able to avoid the generation of vulnerabilities based on its perception of smart contract vulnerabilities.

The dataset of instruction-based smart contract code generation task is $\mathbb{D}=\{(S_1,T_1,C_1),\dots,(S_N,T_N,C_N)\}$, where $S_i$ is a natural language instruction, $T_i$ is a guiding tag and $C_i$ is a smart contract code. Given a natural language instruction $S_i=\{s_1,\dots,s_{|S_i|}\}$, where $s_i$ is a instruction token of $S_i$, and a guiding tag $T_i \in \{security, vulnerable\}$, the goal of CodeBC model is to generate code $C_i=\{c_1,\ldots,c_{|C_i|}\}$ based on the generative probability $P(C_i|S_i,T_i)$ by restricting tag $T_i$ and instruction $S_i$, where $c_i$ is the code token. The $P(C_i|S_i,T_i)$ can be represented as:
\begin{equation}
    P(C_i|S_i,T_i)=\frac{P(C_i)P(T_i|C_i)P(S_i|T_i,C_i)}{P(S_i,T_i)}.
\end{equation}

The overall structure of the CodeBC model consists of three stages, as shown in Figure~\ref{fig:framework}. The first stage is the code infilling task to optimize $P(C_i)$ and the second stage is the vulnerability detection task to optimize $P(T_i|C_i)$. From a probabilistic decomposition perspective, this can theoretically improve the generative probability $P(C_i|S_i,T_i)$. The third stage tags-guided instruction fine-tuning directly optimizes $P(C_i|S_i,T_i)$ based on different $T_i$, namely ``security" and ``vulnerable", which can help the model generate more secure code in the inference process when $T_i=security$.

\subsection{\label{sub}Code Infilling}
Current LLMs can generate smart contract code based on human instructions. However, most of this code fails to compile, showing that these models have poor programming capabilities for Solidity in smart contracts. To equip the model with the ability to write contracts, we set code infilling as the primary training objective.

We adopt the efficient training method from CodeLlama~\citep{codellama}, which splits the text into three parts: prefix, midfix, and suffix. However, smart contracts are often short and typically begin with the version number of the Solidity compiler. Randomly dividing the contract code into three parts, as done in CodeLlama, may prevent the model from effectively learning contextual information. To address this, we adjust the positions of each part, allowing the model to fully utilize contextual information when predicting the midfix code.

As shown on the left side of Figure~\ref{fig:framework}, at this stage, we only use smart contract code as the training data.
Taking a smart contract \( C_i=[c_1, \dots, c_n] \) as an example, we first divide it into five segments of equal length. Random truncation points \( j \) and \( k \) are then selected within the second and fourth segments, respectively, to form \( Pre_i = [c_1, \dots, c_{j-1}] \), \( Mid_i = [c_j, \dots, c_{k-1}] \), and \( Suf_i = [c_k, \dots, c_n] \). Subsequently, the data is randomly organized using either the PSM (Pre-Mid-Suf) or SPM (Suf-Pre-Mid) arrangement, aligning with the work of the OpenAI team on efficiently enhancing the infilling capability of language models~\citep{bavarian2022efficient}.

The three tokens \textless PRE\textgreater , \textless MID\textgreater, and \textless SUF\textgreater are used to denote the beginning of each part.
PSM denotes the combination of prefix, suffix, and midfix,
\begin{equation}\text{\textless PRE\textgreater} {Pre_i} \text{\textless SUF\textgreater} {Suf_i} \text{\textless MID\textgreater} {Mid_i}\end{equation}
SPM denotes the combination of suffix, prefix, and midfix,
\begin{equation}\text{\textless PRE\textgreater} \text{\textless SUF\textgreater} {Suf_i}\text{\textless MID\textgreater} {Pre_i}  {Mid_i}\end{equation}
During the training phase, the \(Mid_i\) in Equations (5) and (6) serves as the target output of the model, while its preceding part as the input. Then we randomly select the combination method among PSM and SPM with probability 0.5.

Given the $i^{th}$ example $C_i=(Pre_i,Mid_i,Suf_i) \in \mathbb{D}$, we minimize the negative log-likelihood of the target $Mid_i$:

{\small
\begin{equation}
    \mathcal{L}_\mathrm{I}= -\sum_{i=1}^{|\mathbb{D}|}\sum_{t=1}^{|Mid_i|}\text{log} \text{P}(m_{t}|m_{1}...m_{t-1},Pre_i,Suf_i;\theta ),
\label{eq:5}
\end{equation}
}
where $m_t$ denotes the token of $Mid_i$, and $\theta$ denotes the parameters of CodeLlama.

\subsection{Vulnerability Detection}\label{sec:vd}
\begin{figure}[!t]
    \centering
    \includegraphics[width=\columnwidth]{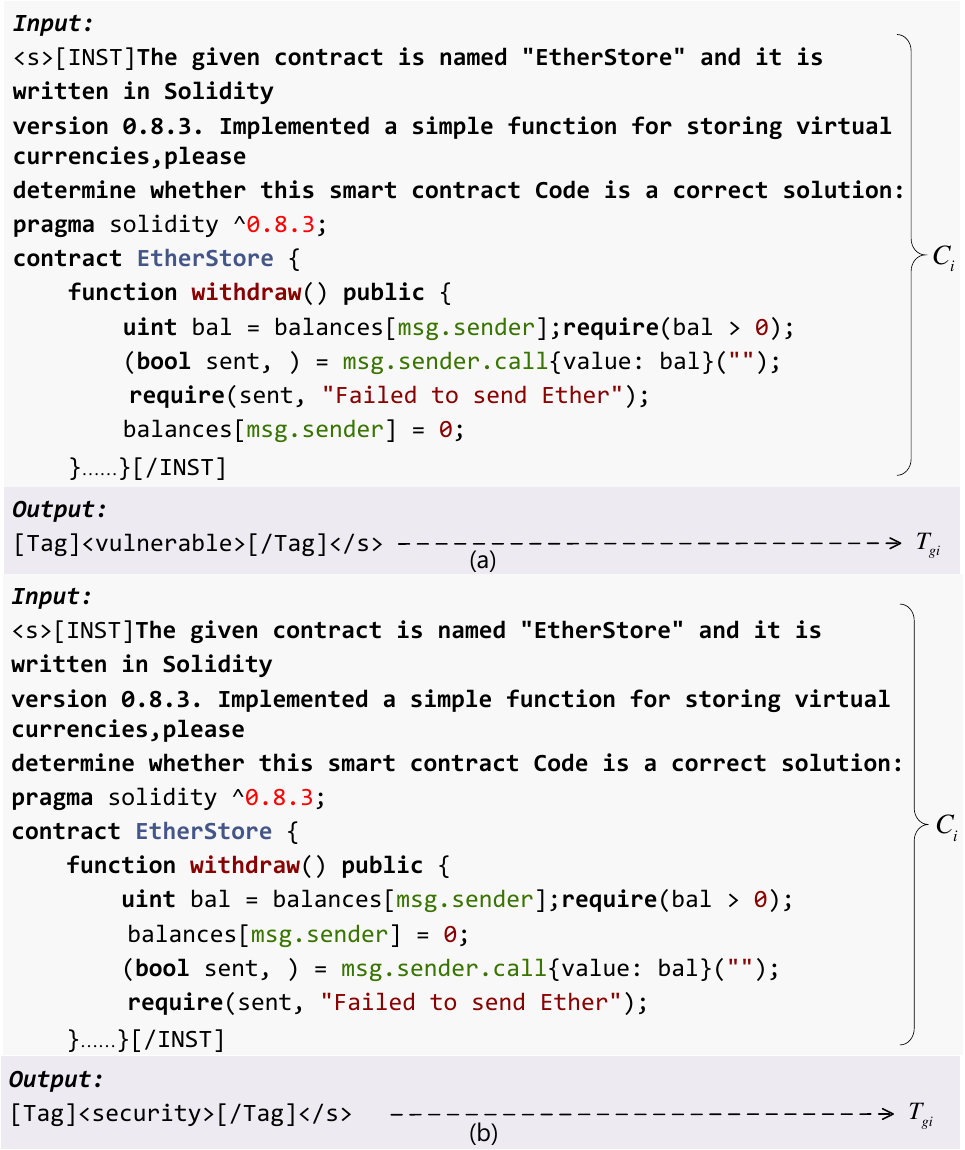}
    \captionsetup{justification=justified, singlelinecheck=false, labelfont=bf}
    \caption{Schematic structure of the training data in the vulnerability detection task. (a) shows the case where the contract code is vulnerable, (b) shows the case where the contract code is not vulnerable}
    \label{fig:detection_struction}
    \vspace{-1em}
\end{figure}

Using a small amount of contract code with vulnerabilities is an efficient and sustainable way to enhance the model's knowledge of contract code security (see experiments in Section~\ref{sub1}). As shown in Figure \ref{fig:detection_struction}, we construct training data by tagging code from open-source vulnerability detection datasets with a ``vulnerable" label and code from smart contract security repositories with a ``security" label. In the second stage, focused on the vulnerability detection task, the model learns to distinguish between contract code with and without vulnerabilities.
The primary goal of the vulnerability detection task is to improve the model's ability to detect security vulnerabilities in smart contracts. This task also provides alignment guidance for the subsequent Tags-guided Instruction Fine-Tuning stage.

Given the code $C_i$, if $C_i$ has no vulnerabilities, the ground-truth output is $T_{g_i} = \{\text{[Tag]}\text{\textless security\textgreater}\text{[/Tag]}\}$, otherwise $T_{g_i} =  \{\text{[Tag]}\text{\textless vulnerable\textgreater}\text{[/Tag]}\}$, where $\text{[Tag]}$, $\text{[/Tag]}$ respectively denote the beginning and the end of the code security detection result.
The model performs vulnerability detection task under a simple prompt ``whether this smart contract Code is a correct solution:" and the negative log-likelihood of the target sentence $T_{g_i}$ is:
\begin{equation}
    \mathcal{L}_\mathrm{D}= -\sum_{i=1}^{|\mathbb{D}|}\sum_{j=1}^{|T_{g_i}|}\text{log} \text{P}(t_{j}|t_{1}...t_{j-1},C_i;\theta_I ),
\label{eq:6}
\end{equation}
where $t_j$ represents the token of $T_{g_i}$, and $\theta_I$ denotes the parameters of equation~\ref{eq:5}.
\subsection{Tags-guided Instruction Fine-tuning}
\begin{figure}[!t]
    \centering
    \includegraphics[width=\columnwidth]{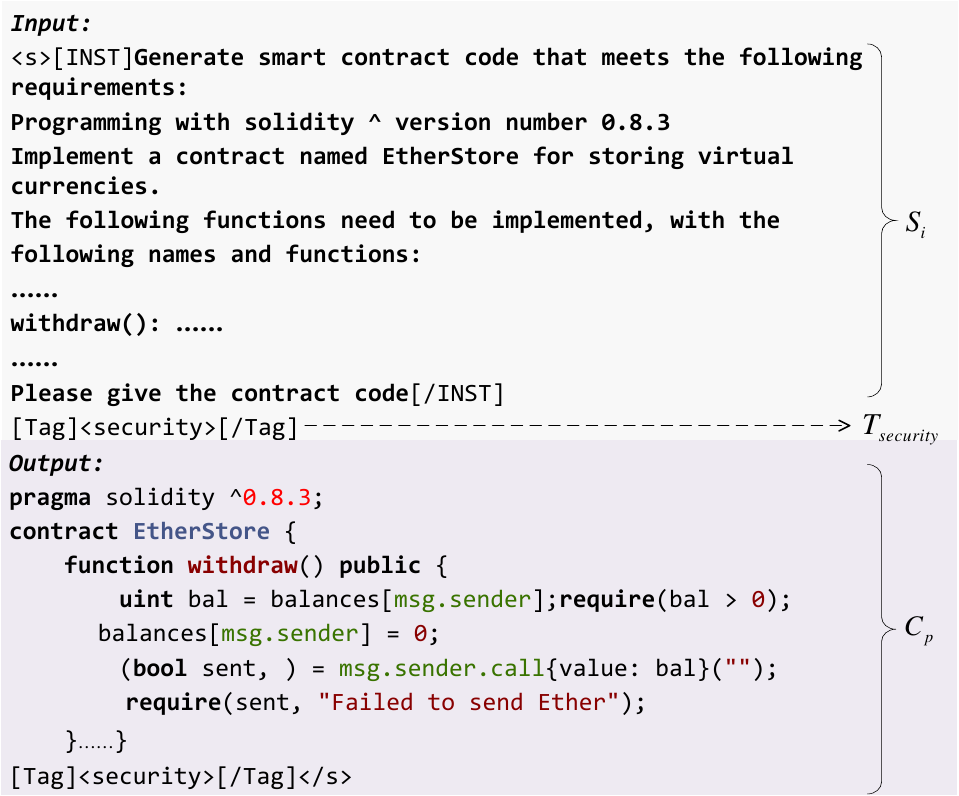}
    \captionsetup{justification=justified, singlelinecheck=false, labelfont=bf}
    \caption{Schematic structure of the training data in the tags-guided instruction fine-tuning}
    \label{fig:tag-guided}
\end{figure}

The training data, built using vulnerability and security tags, helps the language model learn code features more effectively. It also supports the smart contract code generation task by providing tag-based guidance. To align with the vulnerability detection task in the second phase, we design the training data structure for the instruction fine-tuning task, as shown in Figure \ref{fig:tag-guided}. The goal of tags-guided instruction fine-tuning is to use tags to help the model better distinguish between secure and vulnerable code.

In the training process, as shown in the right-bottom of Figure~\ref{fig:framework}, given a human instruction $S_i=[s_1,\dots,s_{|S_i|}]$ and the tag information $T_i \in \{security, vulnerable\}$, the objective of code generation model is to produce the ground-truth code $C_i=[c_1,\dots,c_{|C_i|}]$, where $T_i$ indicates whether the code $C_i$ is security or vulnerable. 
We minimize the negative log-likelihood of the target code:
\begin{equation}
\mathcal{L}_\mathrm{G}= -\sum_{i=1}^{|\mathbb{D}|}\sum_{t=1}^{|C_i|}\text{log} \text{P}(c_{t}|c_{1}...c_{t-1},S_i,\text{PROMPT},T_i;\theta_D ),
\label{eq:7}
\end{equation}
where $c_t$ represents the $t^{th}$ token of $C_i$, and $\theta_D$ denotes the parameters of equation~\ref{eq:6}.

In the inference process, as shown in the right-top of Figure~\ref{fig:framework}, given a human instruction $S_i$ and tag information $T_i=security$, the model will generate safer code $C_p$ with the help of $T_i=security$.

\section{EXPERIMENTAL SETUP}\label{exp setup}
We first introduce some empirical settings, including datasets, evaluation metrics, baselines and parameter settings for CodeBC.

\subsection{Dataset}
\textbf{Training Dataset: SASCsmall-F}\\

The training dataset for the smart contract code generation task is primarily based on the ``Slither Audited Smart Contracts small-multilabel" (SASCsmall) dataset\footnote{https://huggingface.co/datasets/mwritescode/slither-audited-smart-contracts}, which is widely used for vulnerability detection tasks in smart contracts~\citep{tang2023deep, bani2024vulnerability,prifti2024smart}. They utilize Slither~\citep{feist2019slither} to analyze each smart
contract and map the detected vulnerability types to 8 classes
based on the most threatening smart contract vulnerabilities
provided by the Decentralized Application Security Project
(DASP)~\footnote{It is worth noting that our method is not limited to the eight basic vulnerability types of SASCsmall; with appropriate annotated datasets, it can be quickly and efficiently adapted to other types of vulnerabilities.}.
Table~\ref{tabel1} summarizes the dataset composition. It contains 10,571 smart contracts, where 25\% are secure and 75\% are insecure.
To ensure that all smart contracts involved in model training can be deployed in practical application scenarios, we conducted a selection process for smart contracts. We exclude code that encapsulated multiple smart contracts to form complete and complex functionalities from SASCsmall, and finally obtain our dataset SASCsmall-F.

\begin{table}[!t]
\caption{\label{tabel1}The statistics of datasets for smart contract code generation task. Num\_Security and Num\_Insecurity  are the number of security and insecurity smart contracts, respectively.}
\resizebox{\columnwidth}{!}{
\scriptsize
\setlength{\extrarowheight}{3pt}
\begin{tabular}{lcccc}
\toprule
\textbf{Dataset} & \textbf{Total} & \textbf{Num\_Security} & \textbf{Num\_Insecurity}\\
\midrule
SASCsmall & 10,571 & 2,693 & 7,878\\
SASCsmall-F & 7,567 & 1,782 & 5,785\\
SASCsmall-F for CI & 1,782 & 1,782 & 0\\
\bottomrule
\end{tabular}}
\end{table}

Based on SASCsmall-F dataset, we construct different datasets for three fine-tuning stages, and randomly split them into training, validation and test in an 8:1:1 ratio. It is worth noting that the test dataset here is only used in the model analysis experiments to compare whether CodeBC achieves good results on the corresponding tasks. The statistics of these datasets are shown in Table~\ref{tabel1}:
\begin{itemize}
\item Code Infilling dataset(SASCsmall-F for CI): To maximize the model's performance in generating secure code, we only use 1782 secure contract code from the SASCsmall-F for code infilling task. Taking the contract code $[c_1,\dots,c_n]$ as an example, we randomly mask a code segment $Mid_i=[c_j,\dots,c_k-1]$, the prefix $Pre_i=[c_1,\dots,c_{j-1}]$, and the suffix $Suf_i=[c_k,\dots,c_n]$. For the SPM task, LLM use ``$<PRE><SUF>Suf_i<MID>Pre_i$'' as input, and ``$Mid_i$'' as output. For the PSM task, LLM use ``$<PRE>Pre_i<SUF>Suf_i<MID>$'' as input and ``$Mid_i$'' as output. Additionally, each smart contract involved in training is randomly split five times and completely shuffled.
\item Vulnerability Detection dataset: We use all smart contract codes with analysis report results (as security / vulnerability label) from SASCsmall-F dataset for the vulnerability detection task. Taking the contract code $[c_1,\dots,c_n]$ and the target labels $\{security,vulnerable\}$ as examples, the fine-tuning task involved using $[c_1,\dots,c_n]$ as the input. If the contract has no vulnerabilities in analysis reports, the output would be ``$<Tag>security</Tag>$'', otherwise, the the output would be ``$<Tag>vulnerable</Tag>$''.
\item Tags-guided Instruction dataset: We use all smart contract code, code comments(as human instructions) and analysis report results(as security/vulnerability tags) from SASCsmall-F dataset for the tags-guided instruction task. Taking the contract code $C_i$, its comments $S_i$ and the vulnerability labels $\{security,vulnerable\}$ as examples, we design a $\text{PROMPT}$ to generate smart contracts:``Please give the contract code''. As shown in Figure \ref{fig:tag-guided}, when the code has no vulnerabilities, the input is ``$S_i+\text{PROMPT}+\text{[Tag]}\text{\textless security\textgreater}\text{[/Tag]}$'' and the output is $C_i$. When the code has vulnerabilities, the input is ``$S_i+\text{PROMPT}+\text{[Tag]}\text{\textless vulnerable\textgreater}\text{[/Tag]}$'' and the output is $C_i$.
\end{itemize}

\begin{flushleft}
\textbf{Evaluation Dataset: Blockchain-HumanEval}
\end{flushleft}
We construct the Blockchain-HumanEval dataset using the open-source smart contract code repository OpenZeppelin\footnote{https://github.com/OpenZeppelin/openzeppelin-contracts}. This dataset is designed to evaluate whether the generated code meets the requirements of human instructions.
The advantages of Openzeppelin are as follows:
\begin{itemize}
\item As the most popular and verified contract code repository, the code can be considered as correct smart contract code without security vulnerabilities.
\item As a code repository that provides security guarantees for smart contracts, it imposes higher requirements on its security which will place high demands on the security of the model-generated code.
\end{itemize}

The Blockchain-HumanEval dataset selects smart contract code from the OpenZeppelin repository, focusing on code without internal references as the target. To annotate each piece of OpenZeppelin code with human instructions, five blockchain engineers are invited to independently write detailed descriptions. These descriptions include the required Solidity version, names of functions, events, and errors, as well as internal parameters and their respective functionalities.
Afterward, a sixth engineer selects the best instruction from the five submissions. In total, the dataset includes 41 tasks, as detailed in \footnote{https://github.com/wanglingxiang1298/CodeBC/tree/master}. Notably, although the LLM may have been exposed to some OpenZeppelin code during training, it has not encountered the human-annotated instructions. This ensures that when evaluating the LLM's code generation ability, the Blockchain-HumanEval dataset eliminates the risk of data leakage.


\subsection{Baselines and Parameter Settings} 
We compare our CodeBC model with three baselines, including CodeLlama(7B-Instruct)~\citep{codellama}, \\CodeGen25(7B-Instruct)~\citep{nijkamp2023codegen2}, and DeepSeek-Coder(6.7b-Instruct)~\citep{deepseek-coder}. These models are all popular open-source models that possess the capability for instruction-based code generation.

In the model training phase, all three stages of training are performed using Low-Rank Adaptation(LoRA)~\citep{hu2022lora} on the A100 GPU, which has been shown to reduce training time significantly while ensuring good results. The model training process can be viewed as updating the model parameters using the loss values obtained as shown in equations \ref{eq:5} to \ref{eq:7}, thereby minimizing the loss. Assuming the original model parameters are \(W_0\) and the training uses a learning rate \(\eta\), the parameters are updated as follows:
\begin{equation}\Delta{W} = W_0-\eta\frac{\partial L}{\partial W_0},\end{equation}
\begin{equation}W_0 = W_0+\Delta{W}.\end{equation}
Therefore, the objective of model training can be seen as finding the \(\Delta{W}\) that minimizes the loss, which for large-scale language pre-trained models with a vast number of parameters, requires significant computational resources. To address this, LoRA employs a low-rank decomposition to represent parameter updates, that is:
\begin{equation}W_0+\Delta W=W_0+BA,\end{equation}
where, \(B\in\mathbb{R}^{d\times r} and A\in\mathbb{R}^{r\times k}\) are two low-rank decomposition matrices, where the low-rank \(r\)\ used for decomposition satisfies \(r\ll min(d,k)\). By employing this method, we can significantly reduce the consumption of computational resources while achieving fine-tuning of the model. The rank of the update matrix (lora-r) is set to 4, the LoRA scaling factor (lora-alpha) is set to 32, and all parts of the attention block are used as target modules. The learning rate is set to 1e-4, and ten rounds are trained on the Code Infilling and Vulnerability Detection tasks, respectively, and one round is trained on the Tags-guided Instruction task.
In the model inference stage, we run all models using a temperature value of 0.2 and a nucleus sampling method with a parameter value of 0.95. 

\begin{table*}[ht]
\centering
\captionsetup{justification=centering, singlelinecheck=false, labelfont=bf,width=\textwidth}
\caption{\label{table2}The metric-based evaluation results on the Blockchain-HumanEval dataset.} 
\resizebox{\textwidth}{!}{
\scriptsize
\setlength{\extrarowheight}{3pt}
\begin{tabular}{p{1.9cm}ccccccc}
\toprule
\textbf{Models} & \textbf{AvgBLEU} & \textbf{BestBLEU} & \textbf{AvgCB} & \textbf{BestCB} & \textbf{ComPass(\%)} & \textbf{VulRate(\%)} & \textbf{SafeAval(\%)} \\
\midrule
\textbf{CodeGen25} & 0.4298 & 0.5614 & 0.4411 & 0.5111 & 24.87 & 79.02 & 22.43\\
\textbf{DeepSeek-coder} & 0.4251 & 0.5650 & 0.4371 & 0.5084 & 24.39 & 78.04 & 21.95\\
\textbf{GPT-3.5-turbo} & 0.4323 & 0.5160 & 0.6197 & 0.6667 & 82.93 & 39.02 & 60.98\\
\quad+ COT & 0.4507 & 0.5350 & 0.6186 & 0.6763 & 84.88 & 38.05 & 61.95\\
\textbf{GPT-4} & 0.4149 & 0.4851 & 0.6207 & 0.6685 & 84.88 & 34.63 & 65.37\\
\quad+ COT & 0.4040 & 0.4660 & 0.6173 & 0.6667 & 82.44 & 36.10 & 63.90\\
\textbf{CodeLlama} & 0.5099 & 0.5985 & 0.5317 & 0.5816 & 43.90 & 61.95 & 40.48\\ \hline
\textbf{CodeBC(our)} & \textbf{0.6753} & \textbf{0.7288} & \textbf{0.6452} & \textbf{0.6674} & \textbf{86.82} & \textbf{26.34} & \textbf{78.56}\\
\bottomrule
\end{tabular}}
\end{table*}

\begin{table*}[ht]
\centering
\captionsetup{justification=centering, singlelinecheck=false, labelfont=bf,width=\textwidth}
\caption{\label{table3}Ablation results of every stage of CodeBC on the Blockchain-HumanEval dataset.} 
\resizebox{\textwidth}{!}{
\scriptsize
\setlength{\extrarowheight}{3pt}
\begin{tabular}{p{1.9cm}ccccccc}
\toprule
\textbf{Models} & \textbf{AvgBLEU} & \textbf{BestBLEU} & \textbf{AvgCB} & \textbf{BestCB} & \textbf{ComPass(\%)} & \textbf{VulRate(\%)} & \textbf{SafeAval(\%)} \\
\midrule
CodeLlama & 0.5099 & 0.5985 & 0.5317 & 0.5816 & 43.90 & 61.95 & 40.48\\ 
CodeBC & \textbf{0.6753} & \textbf{0.7288} & \textbf{0.6452} & \textbf{0.6674} & \textbf{86.82} & \textbf{26.34} & \textbf{78.56} \\ \hline
CodeBC-CI & 0.5279 & 0.6271 & 0.5739 & 0.6247 & 51.70 & 57.07 & 47.80\\
CodeBC-VD & 0.5759 & 0.6735 & 0.5254 & 0.5633 & 52.68 & 53.66 & 65.33\\
CodeBC-TI & 0.5698 & 0.6298 & 0.6001 & 0.6345 & 77.06 & 44.87 & 68.78\\
\bottomrule
\end{tabular}}
\end{table*}

\subsection{Metrics} 
Each model generates five samples for each natural language instruction from the Blockchain-HumanEval dataset. We evaluate the models based on two aspects: code generation quality and security. The specific evaluation metrics are as follows:

The instruction-based code generation task can also be regarded as a special kind of translation task from natural language to code, therefore, BLEU~\citep{papineni2002bleu} and CodeBLEU~\citep{ren2020codebleu} are employed to evaluate the quality of smart contract code in meeting the requirements of human instructions~\citep{ahmad2023summarize}. We calculate the average BLEU(AvgBLEU), the best BLEU(BestBLEU), the average CodeBLEU(AvgCB) and the best CodeBLEU(BestCB) for evaluation. We do not use the metric pass@k because of the nature of smart contracts that can only be called passively. To better assess the quality of the generated code, we removed all comments and code description sections from the target code and the generated samples, and calculate the metrics only on the code.

The performance of security is evaluated using the review results of the Slither inspection tool~\citep{feist2019slither}, which is the most popular review tool in the Blockchain community and is officially recommended by Ethernet.~\footnote{https://ethereum.org/zh/developers/docs/smart-contracts/testing} The tool returns the compilation results of the code and returns vulnerability analysis results for the compilable contract code. We use the compilation pass rate(ComPass) and the vulnerability rate(VulRate) of the generated code to visualize the performance of the generated results in terms of security, and the safe-availability rate(SafeAval) refers to the proportion of smart contract code that is compilable and free of security vulnerabilities to further reflect the model's ability in terms of security.




\section{EXPERIMENTAL RESULTS}\label{exp result}
In this section, we demonstrate our experiment results and ablation results on BlockChian-HumanEval datasets.

\subsection{Metric-based Evaluation}

\begin{table}[!t]
\captionsetup{justification=justified, singlelinecheck=false, labelfont=bf}
\caption{\label{table4} The test results of CodeLlama and CodeBC-CI model for code infilling task on SASCsmall-F for CI test data.} 
\resizebox{\columnwidth}{!}{
\setlength{\extrarowheight}{3pt}
\begin{tabular}{lcccc}
\toprule
\textbf{Models} & \textbf{AvgBLEU} & \textbf{BestBLEU} & \textbf{AvgCB} & \textbf{BestCB} \\
\midrule
CodeLlama & 0.1842 & 0.1843 & 0.5225 & 0.5231\\\textbf{}
CodeBC-CI & 0.4771 & 0.4829 & 0.8032 & 0.8042\\
\bottomrule
\end{tabular}}
\end{table}
\begin{table*}[!t]
\centering
\captionsetup{justification=centering, singlelinecheck=false, labelfont=bf,width=\textwidth}
\caption{\label{table5}The test results of CodeBC w.o.TI and CodeBC model for vulnerability detection task on SmartBugs dataset.} 
\setlength{\extrarowheight}{3pt}
\begin{tabular}{lccccccccccc}
\toprule
\multirow{2}{*}{\textbf{Models}} & \multicolumn{10}{c}{\textbf{Accuracy}}\\
\cmidrule(lr){2-11}
& RE & AC & AR & ULLC & DoS & BR & FR & TM  & other & all\\
\midrule
CodeBC w.o.TI & 0.90 & 0.33 & 0.0 & 0.40 & 0.50 & 1.0 & 0.25 & 0.80 & 0.67 & 0.52 &\\
CodeBC & 0.90 & 0.17 & 0.0 & 0.42 & 0.67 & 1.0 & 0.50 & 1.0 & 0.67 & 0.52 &\\
\bottomrule
\end{tabular}
\end{table*}


The metric-based evaluation results are shown in Table~\ref{table2}. The results indicate that CodeGen25 and DeepSeek-Coder perform similarly in generation quality. Both are slightly worse than CodeLlama. Their scores on AvgBLEU, BestBLEU, AvgCB, and BestCB are lower than CodeLlama by approximately 15.69\%, 6.67\%, 16.98\%, and 12.07\%, respectively. In terms of security, CodeGen25 and DeepSeek-Coder also perform roughly the same, and are significantly worse than the CodeLlama model. Specifically, they have a 45.45\% lower compilation pass rate, a 27.42\% higher vulnerability rate, and 46.34\% fewer contract codes that are safe to use. These results demonstrate that CodeLlama outperforms both baselines in generation quality and security.

Our CodeBC model performs excellently across all metrics compared to the baselines. In code generation quality, it achieves scores above 0.64 in AvgBLEU, BestBLEU, AvgCB, and BestCB. Compared to the best-performing CodeLlama model, CodeBC shows improvements of approximately 32.43\%, 21.77\%, 21.35\%, and 14.75\% in these metrics, respectively. In terms of security, CodeBC also outperforms CodeLlama. The compilation pass rate increased from 43.90\% to 86.82\%, an increase of 97.77\%. The vulnerability rate decreased from 61.95\% to 26.34\%, a 57.48\% reduction. The percentage of contract code that can be used safely increased from 40.48\% to 78.56\%, a 94.07\% improvement. In conclusion, the improvements of CodeBC over the baseline models demonstrate the effectiveness of our three-stage fine-tuning strategy.
Furthermore, comparative testing with closed-source GPT-series models reveals that while these models demonstrate excellence in code security, CodeBC maintains a performance advantage in benchmark evaluations. This superiority persists even when GPT models are enhanced with explicit chain-of-thought prompting ("Let's think step by step" instructions).

\subsection{\label{sub1}Ablation Experiments}
To verify the actual effect of model training in each stage, we conducted ablation experiments on Blockchain-HumanEval, and the results of each stage are shown in Table \ref{table3}. 
CodeBC-CI denotes the model fine-tuned only through the code infilling task.
CodeBC-VD denotes the model fine-tuned only for the vulnerability detection task.
CodeBC-TI denotes the model fine-tuned only after tag-guided instruction task.

The results show that the training methods in each stage improve the scores of AvgBLEU, BestBLEU, AvgCB, and BestCB. A comparison of CodeLlama, CodeBC-CI, CodeBC-VD, CodeBC-TI, and CodeBC reveals that training across all three stages (CodeBC) contributes to the final model's performance. This suggests that exposing the model to more smart contract code data during training enhances its code generation capability.

In terms of security, the training methods at each stage improved the model's security performance. However, a comparison of CodeBC-CI with other stages shows limited security improvement. Although this stage used only secure contract code for training, the lack of vulnerability knowledge injection limited its impact. This validates the importance of learning vulnerability knowledge. On the other hand, through vulnerability detection task and tag-guided instruction task, CodeBC-VD and CodeBC-TI significantly improved the security of the generated code, as these two tasks helped the model enhance its ability to identify vulnerabilities. The experimental results in Table~\ref{table3} demonstrate that leveraging a small amount of contract code with vulnerabilities is efficient and effective for enhancing the model's knowledge of security vulnerabilities. For instance, training on 7,567 contract vulnerability detection data points (1,782 labeled as secure) reduced the vulnerability rate from 53.66\% to 26.34\%, along with notable improvements in other metrics.

The experimental results show that the efficient code-infilling fine-tuning task enables the model to learn smart contract programming effectively. This significantly enhances CodeBC's ability to generate smart contracts. Meanwhile, the vulnerability detection task and the tags-guided instruction fine-tuning focus on improving the security of model-generated code. These tasks enhance the model's ability to detect vulnerabilities and guide it to generate more secure smart contract code using guiding tags.
\section{DISCUSSIONS}\label{dis}
\begin{figure*}[!h]
    \centering
    \includegraphics[width=\textwidth]{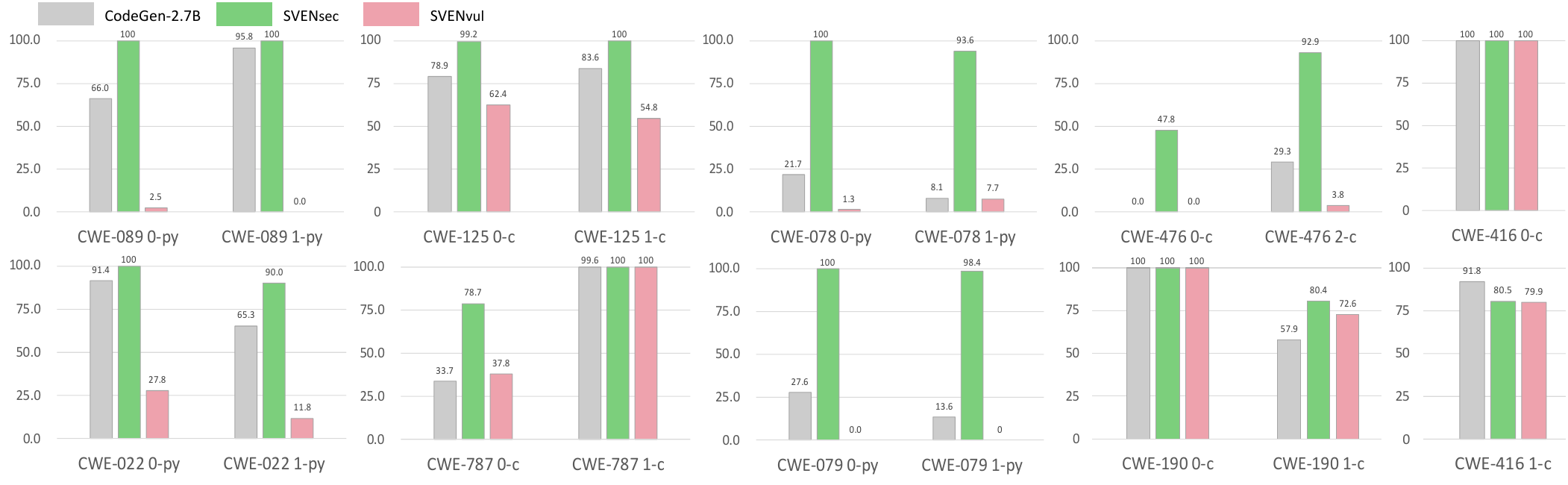}
    \captionsetup{justification=justified, singlelinecheck=false, labelfont=bf}
    \caption{Results of security rate on individual scenarios reported in paper of the SVEN model~\citep{he2023large}. CodeGen-2.7B represents the base model. SVENsec represents the model that maintains secure prefixes, which are expected to guide the model in generating secure code. SVENvul represents the model that maintains vulnerable prefixes, which are expected to guide the model in generating vulnerable code.}
    \label{fig:SVEN}
\end{figure*}
\begin{figure*}[!h]
    \centering
    \includegraphics[width=\textwidth]{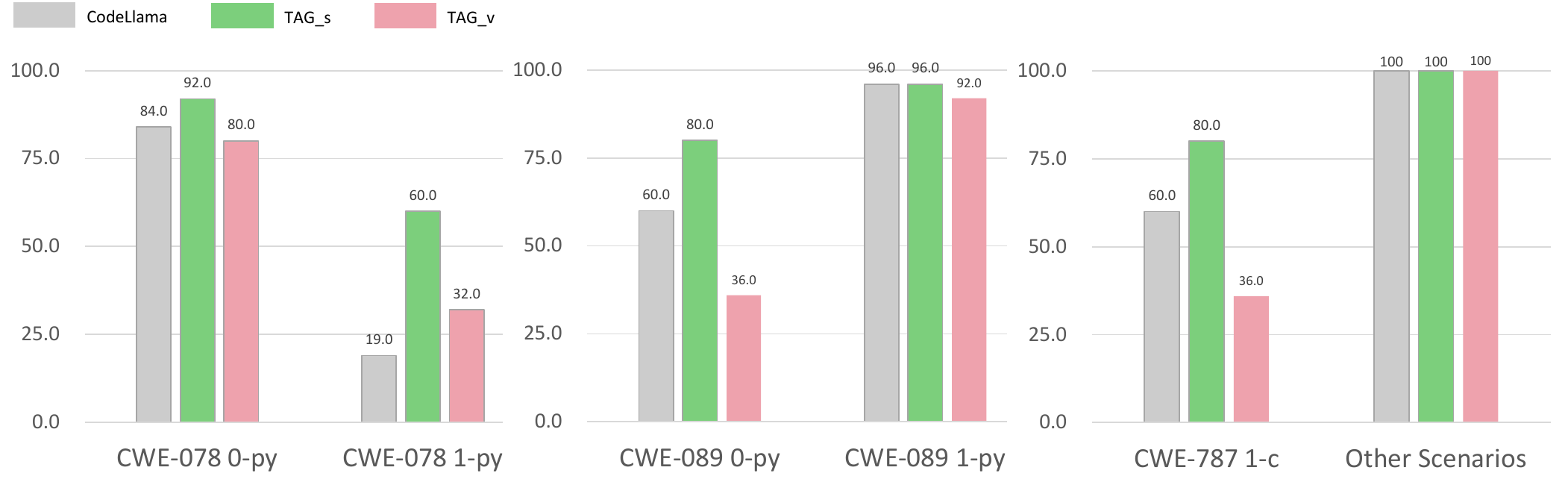}
    \captionsetup{justification=justified, singlelinecheck=false, labelfont=bf}
    \caption{Results of security rate on individual scenarios under method of vulnerability detection. TAG\_s denotes the addition of a $<\text{security}>$ tag to the original prompt. TAG\_v denotes the addition of a $<\text{vulnerable}>$ tag to the original prompt.}
    \label{fig:ourSVEN}
\end{figure*}
In this section, we primarily focus on evaluating the effectiveness of three-stage fine-tuning on various tasks. We test the performance of the CodeBC model on the code infilling task, the vulnerability detection task and other language transferability, respectively, to analyze the reasons for the performance improvement of CodeBC.

\subsection{Code Infilling}
We evaluate the code-infilling capabilities of CodeLlama and CodeBC-CI on the SASCsmall-F for CI test dataset. Table~\ref{table4} presents the results. The CodeBC-CI model achieves better performance than CodeLlama in completing contract code infilling tasks. It shows an improvement of approximately 0.3 in both BLEU and CodeBLEU metrics. This suggests that fine-tuning through code-infilling tasks helps the model learn structural information in smart contract code. It also improves its understanding of code context. As a result, the model enhances its ability to generate smart contracts.

\subsection{Vulnerability Detection}
To verify whether the tag-guided instruction fine-tuning affects the performance of vulnerability detection tasks, handwritten smart contract code from the open-source project SmartBugs~\citep{DurieuxEtAl2020ICSE} is used. This dataset contains 141 smart contract codes with various types of security vulnerabilities. Descriptions of each vulnerability type are available on DASP~\footnote{https://dasp.co/}. We calculate the accuracy of binary classification for vulnerability detection between CodeBC and CodeBC without tag-guided instruction (referred to as CodeBC w.o.TI).
Table \ref{table5} shows the results. The model detects 52\% of the vulnerabilities. Training with tag-guided instruction does not significantly reduce this detection capability. The model demonstrates strong performance in identifying vulnerabilities like Reentrancy (RE) and Bad Randomness (BR). These vulnerabilities often lead to economic losses.
In conclusion, the experiments confirm that our approach can effectively enhance the model's ability to generate secure smart contract codes. We believe the key to improving this capability lies in efficiently incorporating vulnerability-related knowledge without compromising the quality of the generated code. This enables the model to distinguish between secure and insecure code, thereby avoiding vulnerabilities during the code generation process according to the instructions.

\subsection{Transferability Analysis}


In rich-resource programming languages like Python and C/C++, He et al.~\citep{he2023large} proposed the SVEN model for controlled code generation. By comparing vulnerable code with its corresponding fixed code, the SVEN model identifies vulnerability locations more accurately. This improves its vulnerability awareness.
Figure \ref{fig:SVEN} shows the results. The green bar (secure controller) is consistently higher, and the red bar (vulnerable controller) is consistently lower than the gray bar (base model). This demonstrates that the SVEN model significantly enhances the ability to distinguish code security and vulnerabilities. It can generate code that either avoids or includes vulnerabilities based on provided prefixes. However, this method requires precise pairwise vulnerability annotations, which are expensive for low-resource programming languages like Solidity.
To evaluate the transferability of our three-stage fine-tuning method in other languages, we compare the SVEN model with the CodeBC model. Unlike SVEN, CodeBC only uses vulnerability classification tags. We conduct experiments on Python and C/C++ datasets~\citep{he2023large}. Using CodeLlama as the base model, we train it on the SVEN training set with our three-stage fine-tuning method and test it on 18 samples provided by the SVEN model.


Specifically, each sample in the original dataset includes vulnerable code and its corresponding fixed code, with detailed annotations of the fixed locations. We discard the fix location information. Instead, we use pre-fix code snippets as vulnerable negative samples and post-fix code snippets as secure positive samples. These samples are used for multi-stage fine-tuning.
The training and testing sample data, along with the experimental results, will be made publicly available~\footnote{https://github.com/wanglingxiang1298/CodeBC/tree/master}.


The experimental results are shown in Figure~\ref{fig:ourSVEN}. The CodeLlama model itself exhibits exceptional capability in avoiding vulnerabilities during the code generation process. The experimental results in Table~\ref{table2} further confirms that CodeLlama outperforms CodeGen in security performance during code generation. This indicates that the original model performs well in controlled code generation tasks.
Our approach successfully maintains this high level of performance as well. Nevertheless, there are certain situations where CodeLlama does not effectively avoid vulnerabilities. In these situations, the model trained with our method performs similarly to the SVEN model. Specifically, it can generate code that either avoids vulnerabilities or deliberately includes them based on the user's specific requirements. This demonstrates that our method can achieve outcomes comparable to those obtained with extensively annotated datasets, but at a substantially lower cost. This finding validates the transferability and cost-effectiveness of our method.

\section{RELATED WORK}\label{related work}
\subsection{Instruction-based Code Generation}
With the development of LLMs, the usability of code generation models has increased significantly, especially after the successful commercial application of CodeX~\citep{codex}. More and more researchers~\citep{codesys, NL2code,lu-etal-2022-reacc} focus on instruction-based code generation tasks.
AlphaCode~\citep{alphacode} significantly improved the code generation model's ability to understand complex instructions and implement complex codes by generating a large number of code samples and effectively filtering them. However, the expensive execution computation of AlphaCode poses challenges for its real-world application. \citet{nijkamp2022codegen} propose CodeGen, which understands complex instructions by splitting them while controlling model computation costs. Incode~\citep{fried2022incoder} and CodeGeex~\citep{zheng2023codegeex} attempt to enhance the alignment between model-generated code and instructions in different ways. To improve the usability of generated code, \citet{wang-etal-2022-compilable} utilize reinforcement learning to increase the compilation rate of generated code. \citet{zhang-etal-2023-self} patch generated code through error reports and runtime execution results to enhance the correct execution rate of generated code. \citet{yin-etal-2023-natural} propose vertical-domain code generation models for interactive data processing tasks.
However, these works only focus on instruction following, code compilability and execution results, neglecting code vulnerabilities' impact on application stability. 

For code completion tasks( different from our instruction-based code generation task), some researchers~\citep{storhaug2023efficient,he2023large} propose training code generation models using datasets annotated with vulnerability locations to enhance their ability to perceive code vulnerabilities.
\citet{storhaug2023efficient} insert a special token $<\text{VUL CLASS}>$ into the dataset to indicate vulnerability positions and fine-tuned the model on this dataset for code completion task. During inference, when the prediction involves this special token $<\text{VUL CLASS}>$, the output of this token is forcibly altered to prevent the generation of potentially vulnerable code. However, as noted in their limitations~\citep{storhaug2023efficient}, this strict intervention disrupts code fluency and executability, rendering much of the generated code unusable. \citet{he2023large} suggest contrastive learning between the original vulnerable code positions and manually corrected code positions to control the code generation through prefix-tuning for code completion task. However, both these methods are only applicable to popular programming languages with annotated vulnerability position datasets, which cannot be directly applied to the instruction-based smart contract code generation task for blockchain. Therefore, a simpler and more effective method for injecting vulnerability knowledge of smart contracts should be designed to enhance the security of code generation models in the blockchain domain. 

\subsection{Vulnerability Detection}
The task of vulnerability detection is highly crucial in blockchain smart contracts~\citep{brent2018vandal,mehar2019understanding}. 
Oyeente~\citep{luu2016making} is the first smart contract vulnerability detection technique, which detects vulnerabilities by constructing a contract control flow graph, executing symbolic states, and comparing them with properties defined based on vulnerabilities. Subsequently, researchers~\citep{jiang2018contractfuzzer,grishchenko2018semantic,rodler2018sereum} proposed smart contract vulnerability detection techniques based on fuzz testing, taint analysis, and formal verification.
However, these traditional methods have the problems of low coverage and low efficiency. Recently, \citet{tann2018towards} used smart contract data to train a binary classifier based on LSTM. \citet{liu2018s} propose S-gram to identify irregular subsequences as candidate vulnerabilities by training a language model and scanning the token sequences of the target contract for auditing. \citet{he2019learning} propose ILF to use the symbolic execution results as the training dataset for neural networks, and generate input sequences for the test program utilizing neural networks. \citet{gao2020checking} propose SmartEmbed to characterize the stream of contract symbols extracted from AST based on word embedding and vector space techniques, and detect vulnerabilities by comparing the similarity with the vulnerable contracts.
However, current detections rely solely on simple classification models and may not fully comprehend the complex logic of code. In this paper, we design various strategies to enhance vulnerability detection capabilities, thereby preventing the model from generating vulnerable code.

\section{CONCLUSION}\label{conclusion}
In this paper, we propose CodeBC, an instruction-based smart contract code generation model for blockchain. We propose a three-stage fine-tuning strategy to address the security needs of smart contracts. To validate the performance of CodeBC on the smart contract generation task, we propose the Blockchain-HumanEval dataset. Through experiments, we show that our model improves by nearly 32\% in BLEU, 22\% in CodeBLEU, 98\% in compilability rate, 57\% in vulnerability rate, and the percentage of contract code that is safe to use increased from 40\% to 78\%. In the future, we will leverage the control flow graph of smart contracts to design more stringent security policies to ensure that the model can generate secure and usable code.



\printcredits

\bibliographystyle{model1-num-names}

\bibliography{cas-refs}

\subsection*{  } 
\noindent \hspace*{-1mm}
\begin{minipage}{\columnwidth}
\setlength\intextsep{0pt}
\begin{wrapfigure}{l}{25mm}
    \centering
    \includegraphics[width=1in,height=1.25in,clip,keepaspectratio]{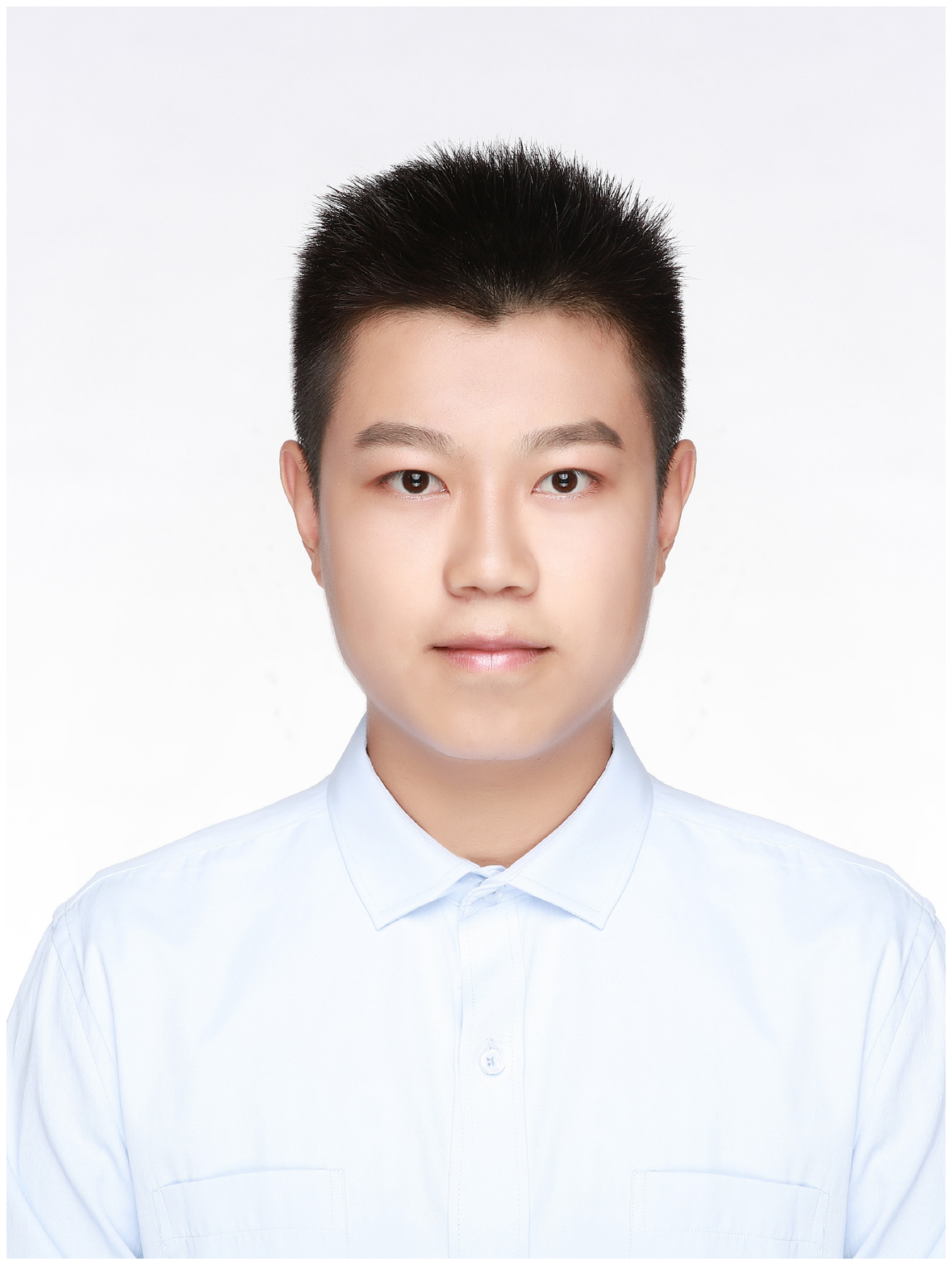}
\end{wrapfigure}
\noindent {\small 
\textbf{Lingxiang Wang} received the Bachelor's degree in Information Security from the School of Control and Computer Engineering, North China Electric Power University, Beijing, China, in 2022. He is currently pursuing his Ph.D. in Artificial Intelligence at the School of Artificial Intelligence, Beihang University. His research interests include natural language processing, code generation and language model}
\end{minipage}\par

\hspace*{\fill}

\subsection*{  } 
\noindent \hspace*{-1mm}
\begin{minipage}{\columnwidth}
\setlength\intextsep{0pt}
\begin{wrapfigure}{l}{25mm}
    \centering
    \includegraphics[width=1in,height=1.25in,clip,keepaspectratio]{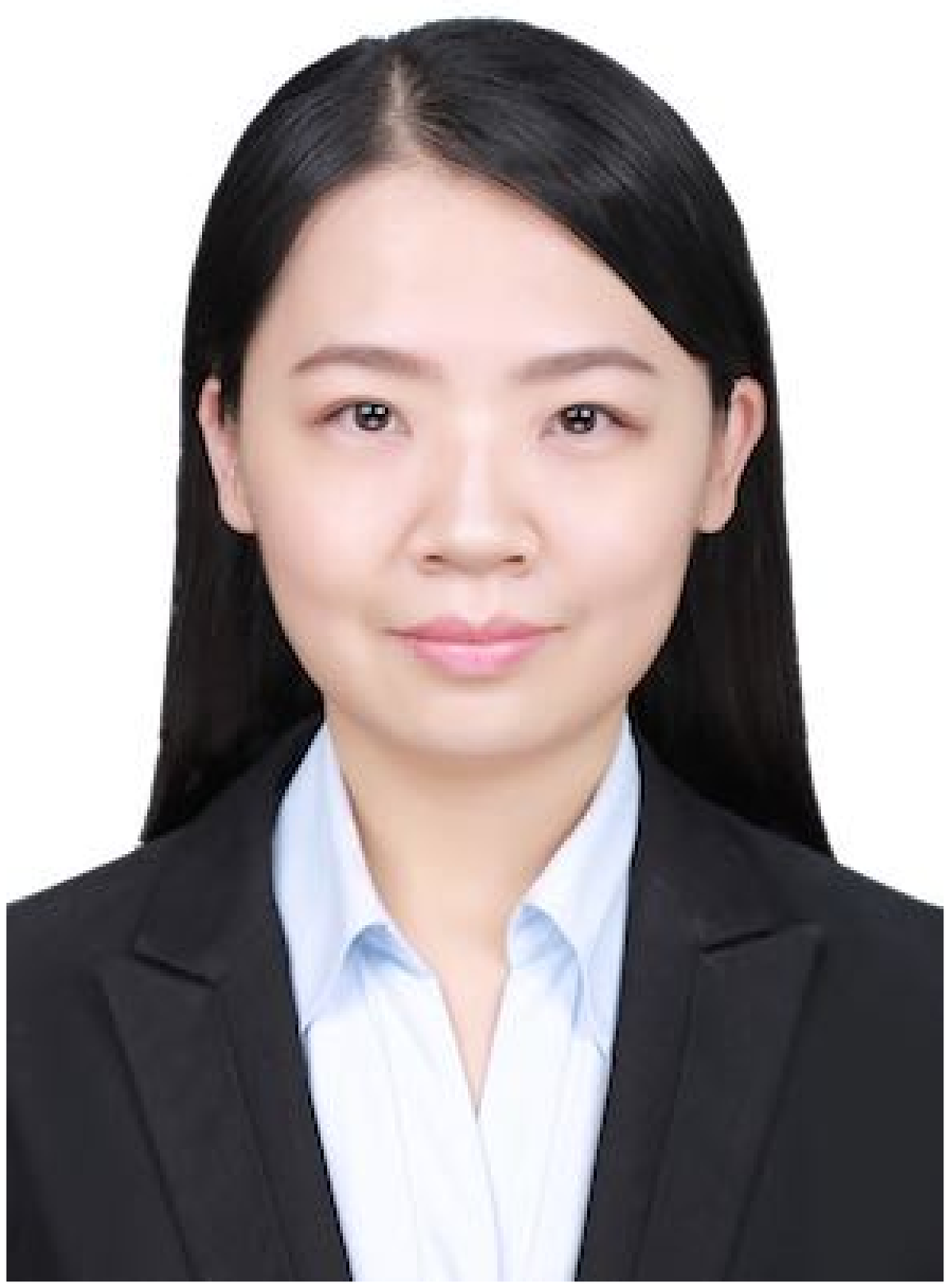}
\end{wrapfigure}
\noindent {\small 
\textbf{Hainan Zhang} (Member, IEEE) received the PhD degree in computer science and engineering from the Institute of Computing and Technology, Chinese Academy of Sciences, in 2019. She is currently a associate researcher with the School of Artificial Intelligence, Beihang University. Her research interests include code generation, language model and dialogue system. She has published several papers on ACL, SIGIR, AAAI etc.}
\end{minipage}\par

\hspace*{\fill}

\subsection*{  } 
\noindent \hspace*{-1mm}
\begin{minipage}{\columnwidth}
\setlength\intextsep{0pt}
\begin{wrapfigure}{l}{25mm}
    \centering
    \includegraphics[width=1in,height=1.25in,clip,keepaspectratio]{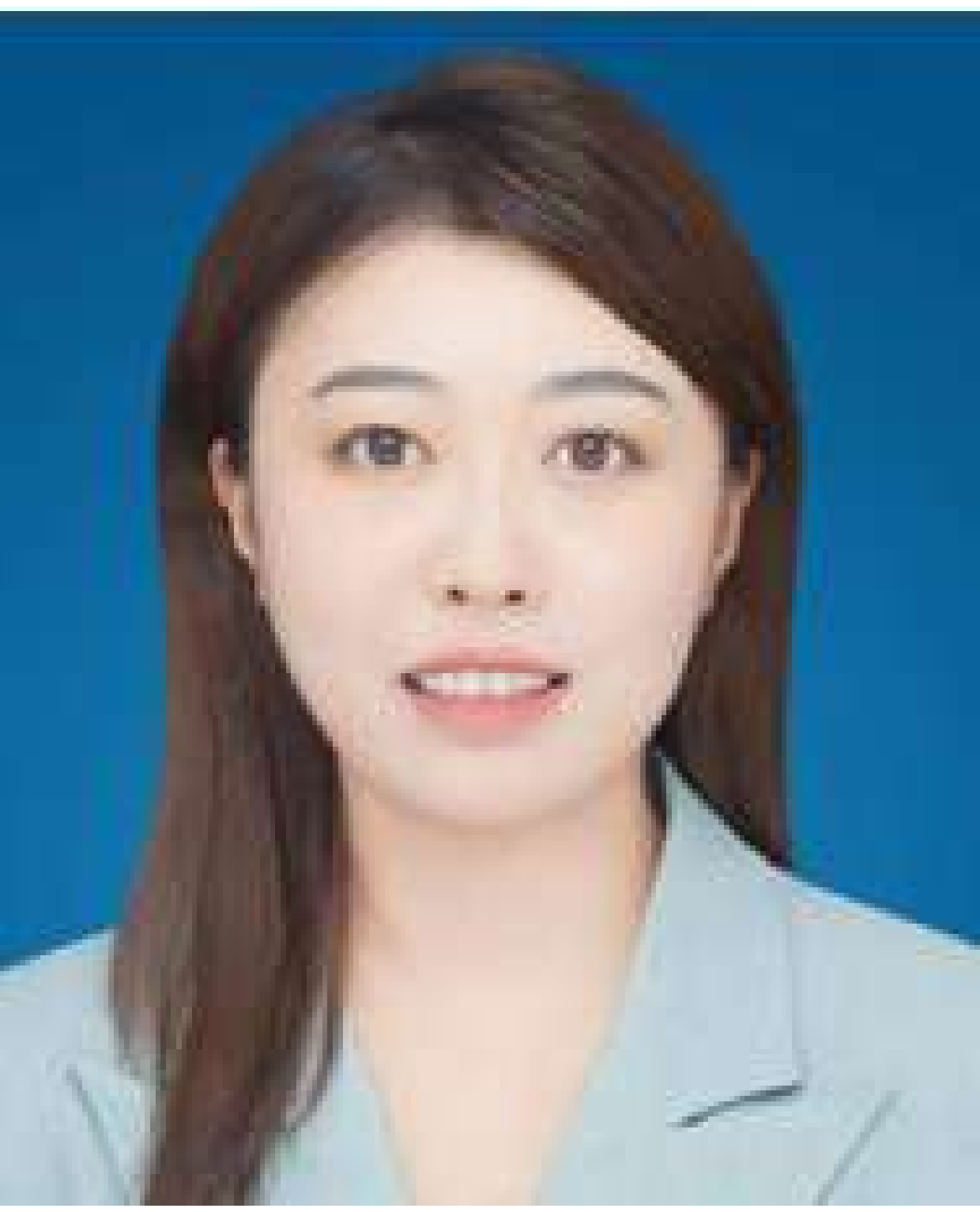}
\end{wrapfigure}
\noindent {\small 
\textbf{Qinnan Zhang} (Student Member, IEEE) received the Ph.D. degree from the Central University of Finance and Economics, Beijing, China, in 2023. She is currently a Postdoctoral Fellow with the Institute of Artificial Intelligence, Beijing Advanced Innovation Center for Future Blockchain and Privacy Computing, Beihang University, Beijing. Her current research interests include blockchain, federated learning, incentive mechanism, game theory, and edge intelligence.}
\end{minipage}\par

\hspace*{\fill}
\subsection*{  } 
\noindent \hspace*{-1mm}
\begin{minipage}{\columnwidth}
\setlength\intextsep{0pt}
\begin{wrapfigure}{l}{25mm}
    \centering
    \includegraphics[width=1in,height=1.25in,clip,keepaspectratio]{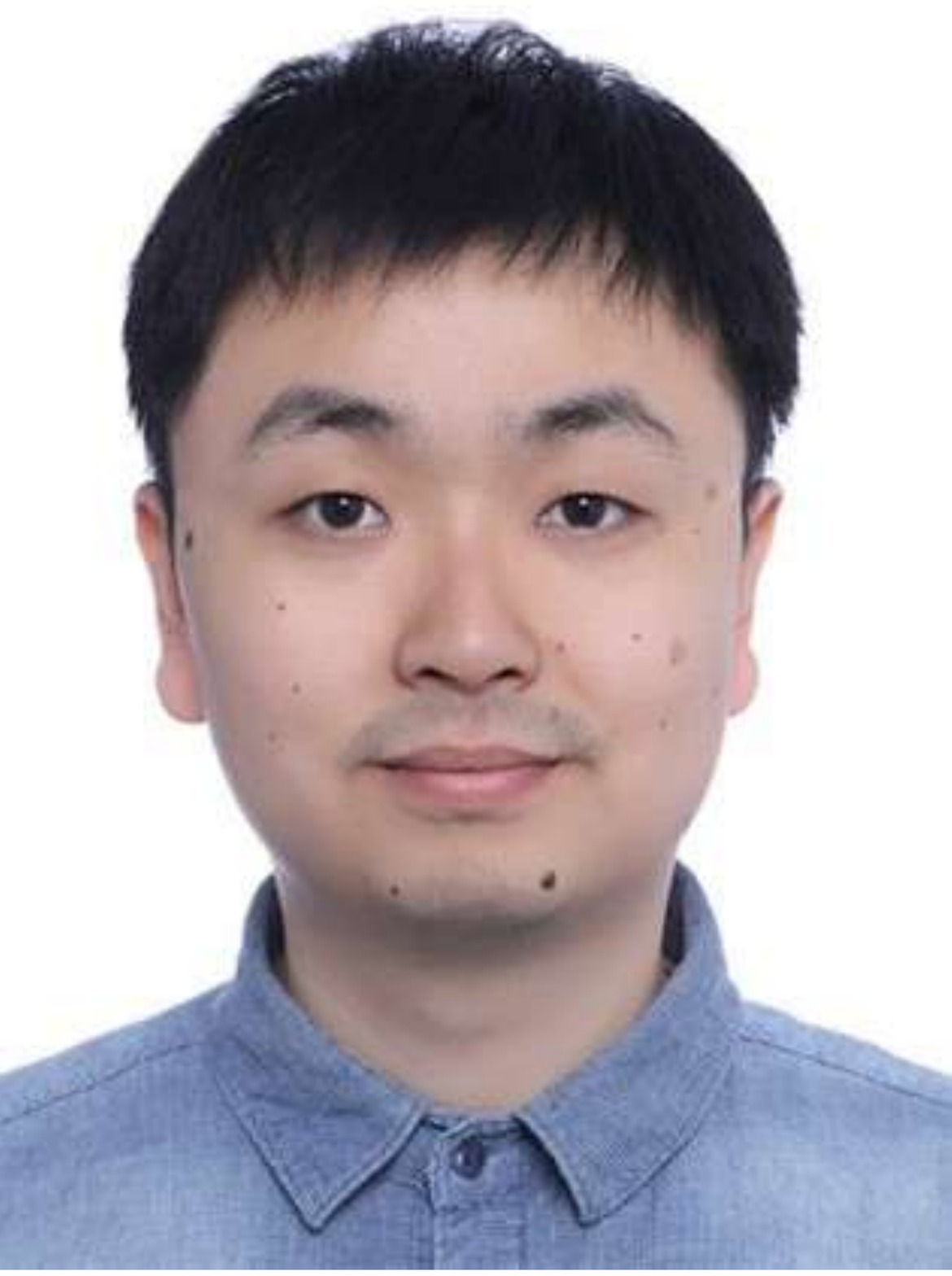}
\end{wrapfigure}
\noindent {\small 
\textbf{Ziwei Wang} received the Ph.D degree in signal processing from Aerospace Information Research Institute, Chinese Academy of Sciences, Beijing, China, in 2016. He is currently an Associate Professor with the School of Artificial Intelligence/ Beijing Advanced Innovation Center for Future Blockchain and Privacy Computing, Beihang University. His current research interests include intelligent signal processing, integrated sensing and communication systems, and game theory.
}
\end{minipage}\par

\hspace*{\fill}

\subsection*{  } 
\noindent \hspace*{-1mm}
\begin{minipage}{\columnwidth}
\setlength\intextsep{0pt}
\begin{wrapfigure}{l}{25mm}
    \centering
    \includegraphics[width=1in,height=1.25in,clip,keepaspectratio]{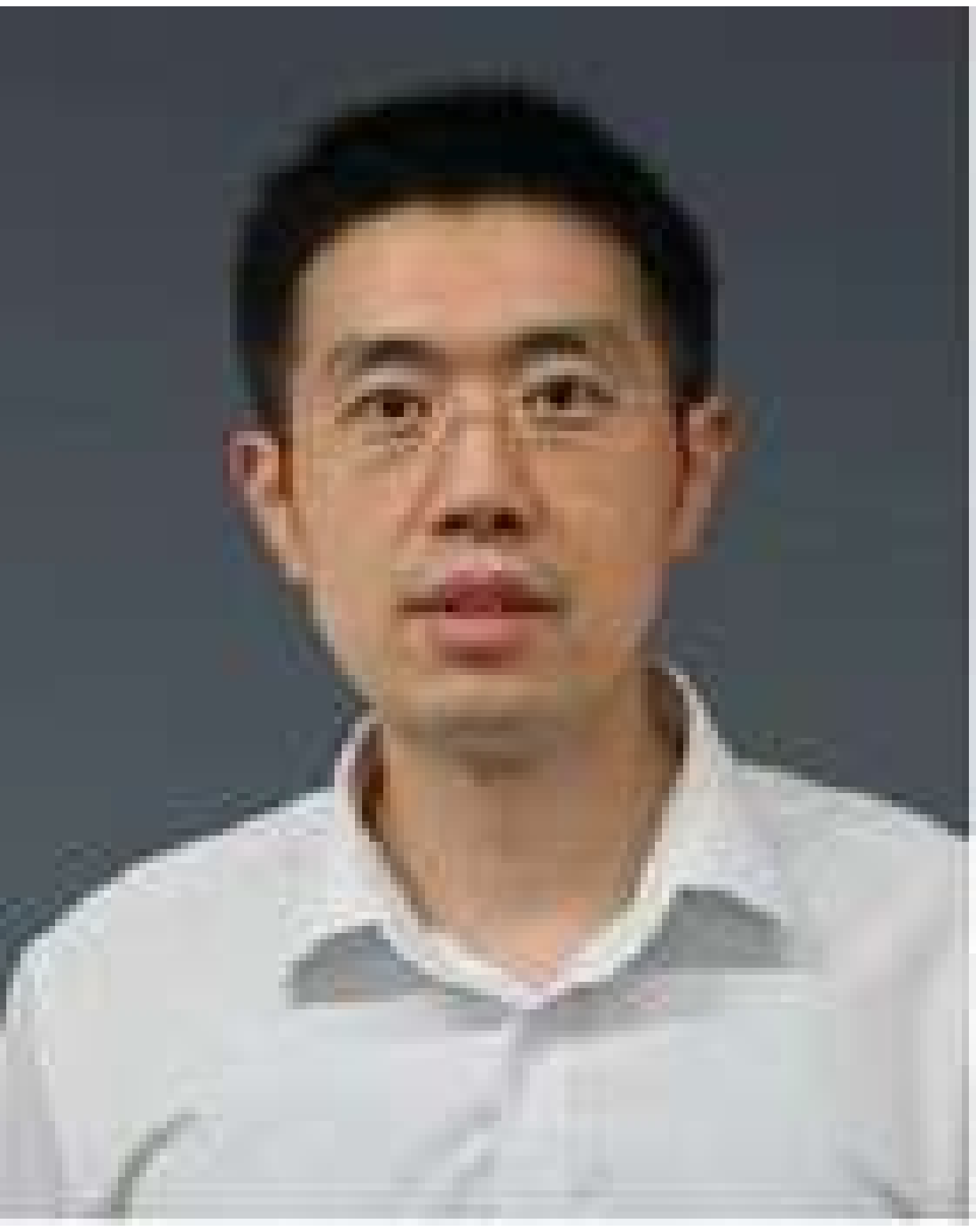}
\end{wrapfigure}
\noindent {\small 
\textbf{Hongwei Zheng} received the Doctoral degree in quantitative economics from The Ohio State University, Columbus, OH, USA, in 2010. He is a Senior Researcher with Beijing Academy of Blockchain and Edge Computing, Beijing, China. His research interests include complex networks, blockchain, and artificial intelligence.}
\end{minipage}\par

\hspace*{\fill}

\subsection*{  } 
\noindent \hspace*{-1mm}
\begin{minipage}{\columnwidth}
\setlength\intextsep{0pt}
\begin{wrapfigure}{l}{25mm}
    \centering
    \includegraphics[width=1in,height=1.25in,clip,keepaspectratio]{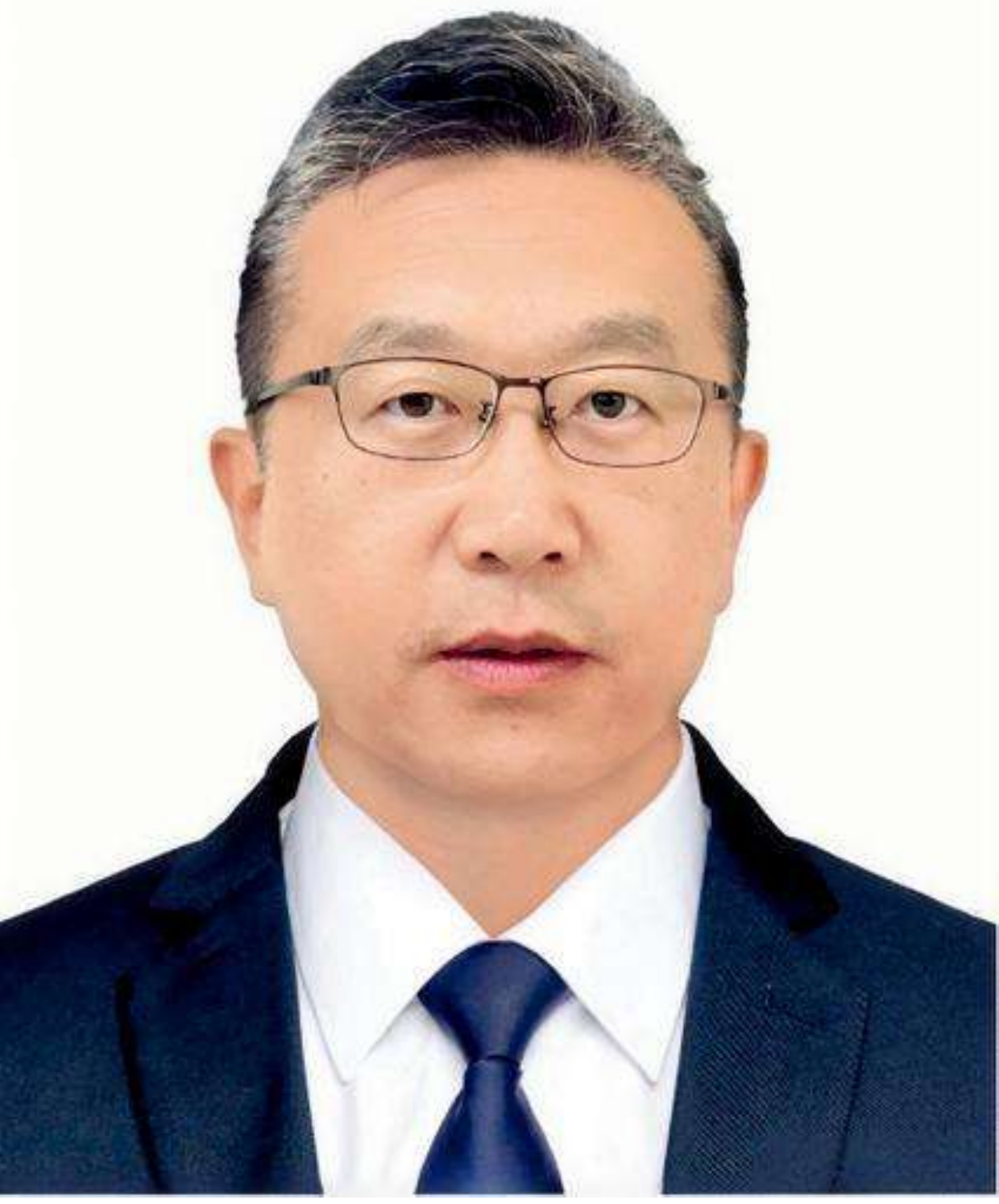}
\end{wrapfigure}
\noindent {\small 
\textbf{Jin Dong} is the General Director of Beijing Academy of Blockchain and Edge Computing. He is also the General Director of Beijing Advanced Innovation Center for Future Blockchain and Privacy Computing. The team he led developed “ChainMaker”, the first hardware-software integrated blockchain system around the globe. He has been long dedicated in the research areas such as blockchain, artificial intelligence and low-power chip design.}
\end{minipage}\par

\hspace*{\fill}

\subsection*{  } 
\noindent \hspace*{-1mm}
\begin{minipage}{\columnwidth}
\setlength\intextsep{0pt}
\begin{wrapfigure}{l}{25mm}
    \centering
    \includegraphics[width=1in,height=1.25in,clip,keepaspectratio]{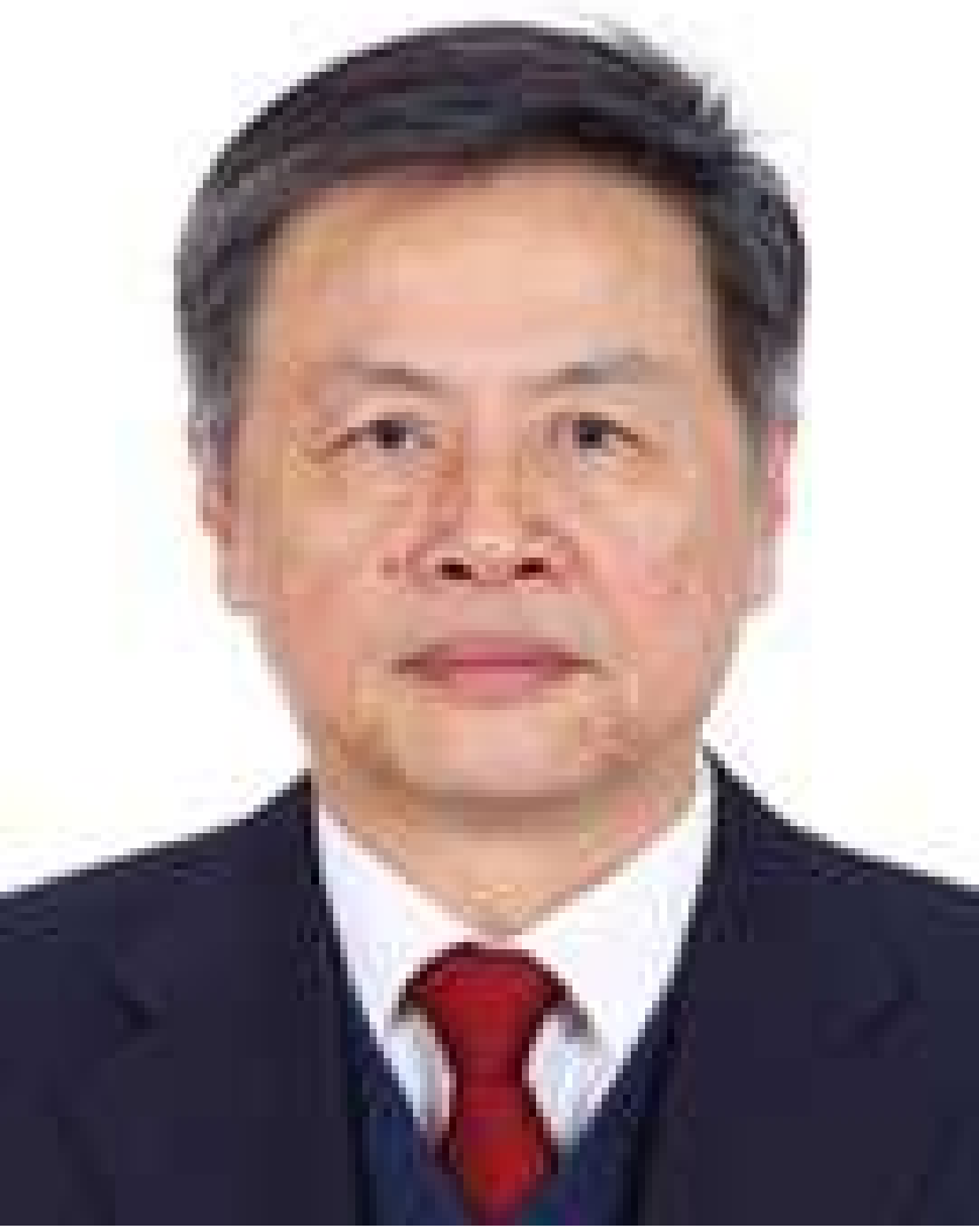}
\end{wrapfigure}
\noindent {\small 
\textbf{Zhiming Zheng} received the Ph.D. degree in mathematics from the School of Mathematical Sciences, Peking University, Beijing, China, in 1987. He is currently a Professor with the Institute of Artificial Intelligence, Beihang University, Beijing, China. He is one of the Initiators of Blockchain ChainMaker. His research interests include refined intelligence, blockchain, and privacy computing. Prof. Zheng is a member of Chinese Academy of Sciences.}
\end{minipage}\par

\end{document}